\documentclass[11pt,twoside]{article}

\usepackage{graphicx} 
\usepackage{mathptmx} 
\usepackage{amsmath}
\usepackage{amssymb}   

\usepackage{subfigure}
\usepackage{psboxit}
\usepackage{epsfig}
\usepackage{times}
\usepackage{wrapfig}
\usepackage{enumerate}
\usepackage{lscape}
\usepackage{longtable}
\usepackage{moreverb}
\usepackage{latexsym}
\usepackage{rotating}
\usepackage{natbib}
\usepackage[dvips]{color}

\def\farcm{\hbox{$.\mkern-4mu^\prime$}}
\def\aap{A\&A}%
\def\aaps{A\&AS}%
\def\aj{AJ}%
\def\apj{ApJ}%
\def\apjs{ApJS}%
\def\mnras{MNRAS}%
\def\pasp{PASP}%

\begin{document}

\markboth{C. Papadaki et al. }{Photometric study of selected cataclysmic variables II.}
\title{\parbox{\textwidth}{ {\normalsize \sl 2008: The Journal of Astronomical Data 14, 1.    \hfill
\copyright\ C. Papadaki et al. }}\\ 
{\vspace{5mm} Photometric study of selected cataclysmic variables II.\\ 
Time-series photometry of  nine  systems. }
}
\author{C. Papadaki$^{1,2}$, H.M.J. Boffin$^{3}$,  V. Stanishev$^{4}$, P. Boumis$^{5}$, \\  {\smallskip}
 S. Akras$^{5,6}$ and  C. Sterken$^{1}$        \\  \vspace{3mm}
\\
\small (1) Vrije Universiteit Brussel, Pleinlaan 2, 1050 Brussels, Belgium\\ 
\small (2) ESO, Casilla 19001, Santiago19, Chile \\
\small (3) ESO, Karl-Schwarzschild-Str. 2, 85748 Garching, Germany \\
\small (4) Department of Physics, Stockholm University,  106 91 Stockholm, Sweden \\
\small (5) Institute of Astronomy and Astrophysics, National Observatory of Athens,  15236 Athens, Greece\\
\small (6) University of Crete, Physics Department, 71003 Heraklion, Crete, Greece}

\date{\small Received  June 2008, accepted    2008}

\maketitle

\section*{Abstract}
We present time-series photometry of nine cataclysmic variables: EI UMa, V844Her, V751 Cyg, V516 Cyg, GZ Cnc, TY Psc, V1315 Aql, ASAS J002511+1217.12, V1315 Aql and LN UMa. The observations were conducted at various observatories, covering 170  hours and comprising 7,850 data points in total.\\

\noindent For the majority of targets we confirm previously reported periodicities and for some of them we give, for the first time, their spectroscopic orbital periods. For those dwarf-nova systems which we observed during both quiescence and outburst, the increase in brightness was followed by a decrease in the amount of flickering. Quasi-periodic oscillations have either been discovered,  or were confirmed. For the eclipsing system V1315 Aql we have covered 9 eclipses, and obtained a refined orbital ephemeris.  We find that, during its long baseline of observations, no change in the orbital period of this system has occurred. V1315 Aql also shows eclipses of variable depth.

\section{Introduction}             

Cataclysmic variables (CVs) consist of a low-mass star filling its Roche lobe and transferring mass to its white-dwarf (WD) companion, a procedure resulting in the formation of an accretion disc. Due to their unpredictability and diversity leading to various types of CVs, these stars are since long    the subject of many studies. 

Our aim is to better understand the properties of these systems by detecting periodicities and by analysing the various physical processes that occur, such as outbursts, superhumps, eclipses, flickering, etc. Our data will be useful for studying these CVs in the future, when a system possibly is in a different brightness state. In this study we present time-resolved photometric data of nine poorly-studied CVs, dwarf novae (DN) and nova-likes (NLs) in particular. A deeper study of five more CVs, was published by \citet{pap06}.

\section{Observations}

During the period 2002--2005 we conducted observations of several targets at five different observatories:  the South African Astronomical Observatory (SAAO), the Observatorium Hoher List in Germany (OHL), Uccle Observatory (UC) in Belgium and two observatories in Greece: Kryoneri (KR) and Skinakas (SK). At the SAAO we used the 1-m telescope equipped with a back-illuminated $1024\times1024$, 24-micron pixel STE4 CCD camera with a field of view (FOV) of $5\farcm3 \times 5\farcm3$. At the OHL, we used the 1-m Cassegrain-Nasmyth telescope with the $2048\times2048$, 15-micron pixel HoLiCam CCD camera with a 2-sided read-out. We used the focal reducer and an effective FOV of $14\farcm1 \times 14\farcm1$. At UC we used the 0.85-m Schmidt telescope with a $2048\times2048$, 9-micron pixel Princeton Instruments TE CDD camera with a $30\farcm1 \times 30\farcm1$ FOV restricted to  an effective FOV of $6\farcm1 \times 6\farcm1$ on average. At KR the 1.2-m Cassegrain telescope with a $512 \times 512$ 24-micron pixel CCD camera and a $2\farcm5 \times 2\farcm5$ FOV. Finally, we performed observations with the 1.3-m Ritchey-Chr\'eti{e}n telescope at SK, giving a FOV of $8\farcm5 \times 8\farcm5$ and used the 1024x1024 pixel SITe CCD camera.

While most of the observations were unfiltered, the Johnson-$R$ filter was occasionally used at SK. Table~\ref{t_log_all} gives the complete observing log for our 9 targets. Whenever a filter was used it is indicated as a superscript in the UT date. The heliocentric Julian date (HJD) at the beginning of each  observing run   as well as the total duration in hours, are also given. Fig.~\ref{f_fov_chart1} gives the finding charts along with the selected comparison stars.

\begin{table}
\caption{Log of observations.}
\label{t_log_all}
\centering
\begin{tabular}{|llll|llll|}
\hline
UT date & Site & $\rm HJD_{\rm start}$ & Dur.& UT date & Site & $\rm HJD_{\rm start}$ & Dur.\\
\hline  
 {\it EI UMa}  & &	  &     &                {\it GZ Cnc}& & 	  &     \\
 09dec02 & OHL & 2618.424 & 2.6 &		 09mar03 & OHL & 2708.394 & 2.8 \\
 10dec02 & OHL & 2619.417 & 7.7 &		 16mar03 & UC  & 2715.451 & 0.7 \\
 31jan03 & OHL & 2671.466 & 4.7 &		 22mar03 & UC  & 2721.369 & 0.7 \\
 31jan03 & UC  & 2671.459 & 5.0 &		 25mar03 & UC  & 2724.430 & 1.7 \\
 14feb03 & UC  & 2685.461 & 4.8 &		         &     &          &     \\
 16feb03 & UC  & 2687.315 & 6.3 &                {\it TY Psc}& &    &	 \\	 
 17feb03 & UC  & 2688.320 & 8.6 &		 21jul03 & SK  & 2842.526 & 0.3\\
       	 &     & 	  &     & 	 	 22jul03 & SK  & 2843.523 & 1.2\\
 {\it V844 Her}& &	   &   &		 23jul03 & SK  & 2844.491 & 1.9\\
 04jul01 & UC  & 2095.423 & 3.2 &		 10jan05 & OHL & 3381.255 & 1.9\\
 15may02 & UC  & 2410.497 & 2.0 &                        &     &          &    \\
 16may02 & UC  & 2411.379 & 5.4 &		{\it ASAS J0025+1217.12}  & &  & \\
 12jun02 & SK  & 2438.282 & 2.0 &		07oct04 & OHL  & 3286.333   & 1.7\\
 14jun02 & SK  & 2440.282 & 2.5 &		10oct04 & OHL  & 3289.498   & 2.0\\
 15jun02 & SK  & 2441.282 & 2.5 &		11oct04 & OHL  & 3290.423   & 1.8\\
	 &     & 	  &     &	                &      &            &    \\    
{\it V751 Cyg} &  &       &     &	  	{\it V1315 Aql}& &	 &     \\	
14may04  & OHL & 3140.446 & 3.7 &		14aug03 & SAAO& 2866.259 & 2.0 \\	
15may04  & OHL & 3141.448 & 0.7 &		15aug03 & SAAO& 2867.260 & 3.5 \\	
16may04  & OHL & 3142.444 & 3.5 &		16aug03 & SAAO& 2868.275 & 1.5 \\       
24may04  & OHL & 3150.529 & 1.5 &		17aug03 & SAAO& 2869.224 & 4.2 \\	
26may04  & OHL & 3152.411 & 4.0 &		22aug04 & KR  & 3240.287 & 3.7 \\	
27may04  & OHL & 3153.411 & 4.0 &		31aug04$^{R}$ & SK  & 3249.242 & 5.4 \\  
28may04  & OHL & 3154.407 & 4.3 &  		01sep04$^{R}$ & SK  & 3250.254 & 5.2 \\  
29may04  & OHL & 3155.410 & 2.5 &		02sep04$^{R}$ & SK  & 3251.267 & 4.5 \\  
07may05  & KR  & 3498.458 & 3.3 &                       &     &          &     \\
08may05  & KR  & 3499.455 & 3.5 & 		{\it LN UMa}  & & &	 \\	   
09may05  & KR  & 3500.504 & 2.3 &		18feb03  & UC & 2689.452 & 1.7 \\      
13may05  & KR  & 3504.465 & 3.5 &		25feb03  & UC & 2696.288 & 9.2  \\     
	 &     & 	  &     &		01jun04$^{R}$  & SK & 3158.330 & 2.0\\ 
 {\it V516 Cyg}& &	  &     &		02jun04$^{R}$  & SK & 3159.375 & 1.4\\ 
 08oct02 & UC  & 2556.319 & 3.5 &		06jun04$^{R}$  & SK & 3163.281 & 3.0 \\
 09oct02 & UC  & 2557.331 & 2.8 &		        &     &          &     \\
 19oct02 & UC  & 2567.323 & 2.0 &		        &     &          &     \\
 23oct02 & UC  & 2571.297 & 1.4 &		        &     &          &     \\
 05nov02 & UC  & 2584.271 & 3.2 &		        &     &          &     \\
\hline
  \end{tabular}
 \begin{center}
 {\footnotesize 
 	     Notes: $\rm HJD_{\rm start}=\rm HJD-2450000$; Dur. is the duration of each run in hours; if any filter was used, the photometric band is noted as a superscript at the corresponding UT date.\hfill}
 \end{center}
 \end{table}																							  	     
 
 \begin{figure}
   \centerline{\includegraphics[width=0.95\textwidth]{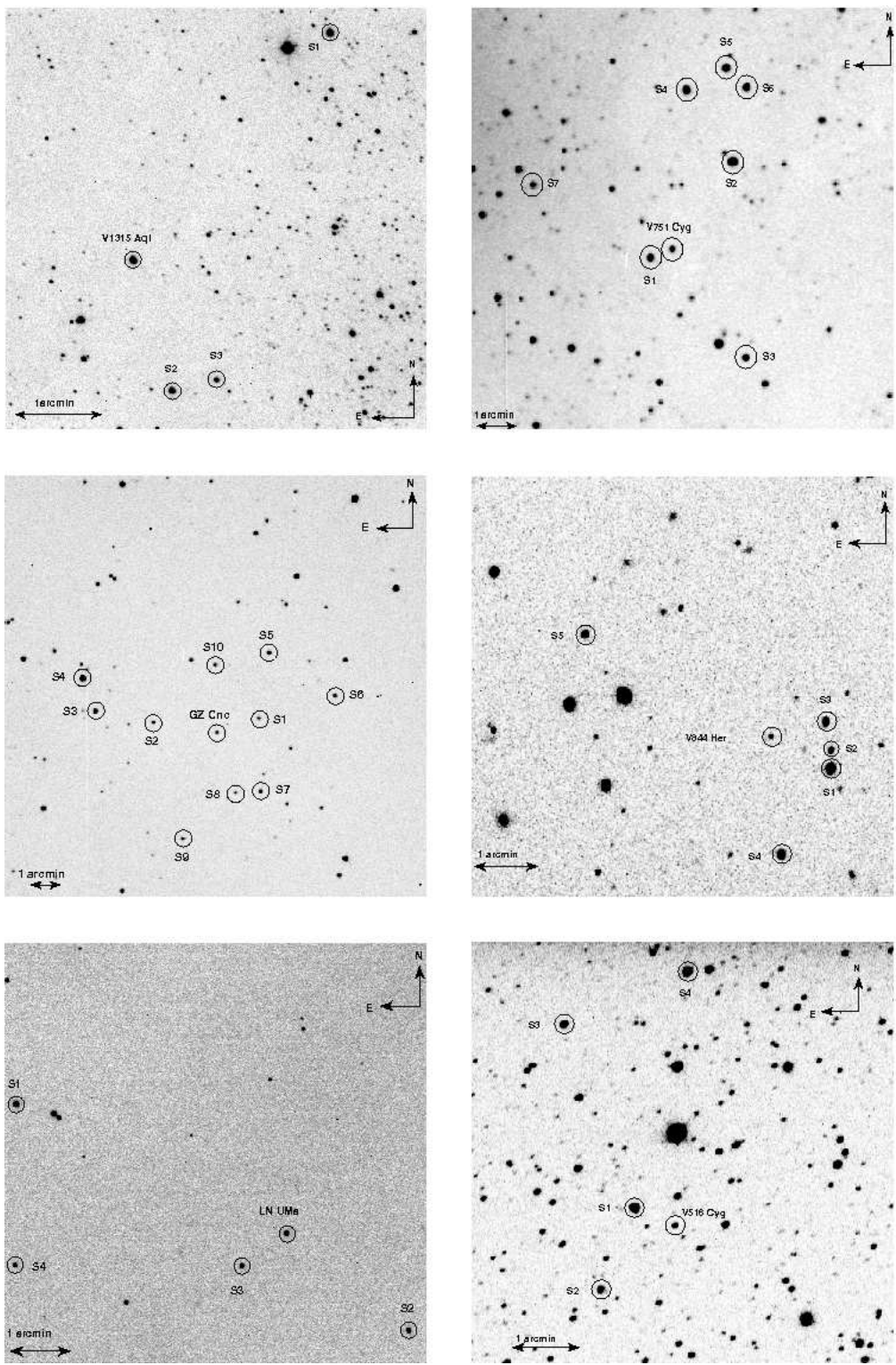}}
       \caption{The finding charts of V1315 Aql, V751 Cyg, GZ Cnc, V844 Her, LN UMa and V516 Cyg. Marked on each chart are the comparison stars that served for differential photometry.}
       \label{f_fov_chart1}
     \end{figure}
 
\setcounter{figure}{0}

\begin{figure}[!ht]
  \centerline{\includegraphics[width=1\textwidth]{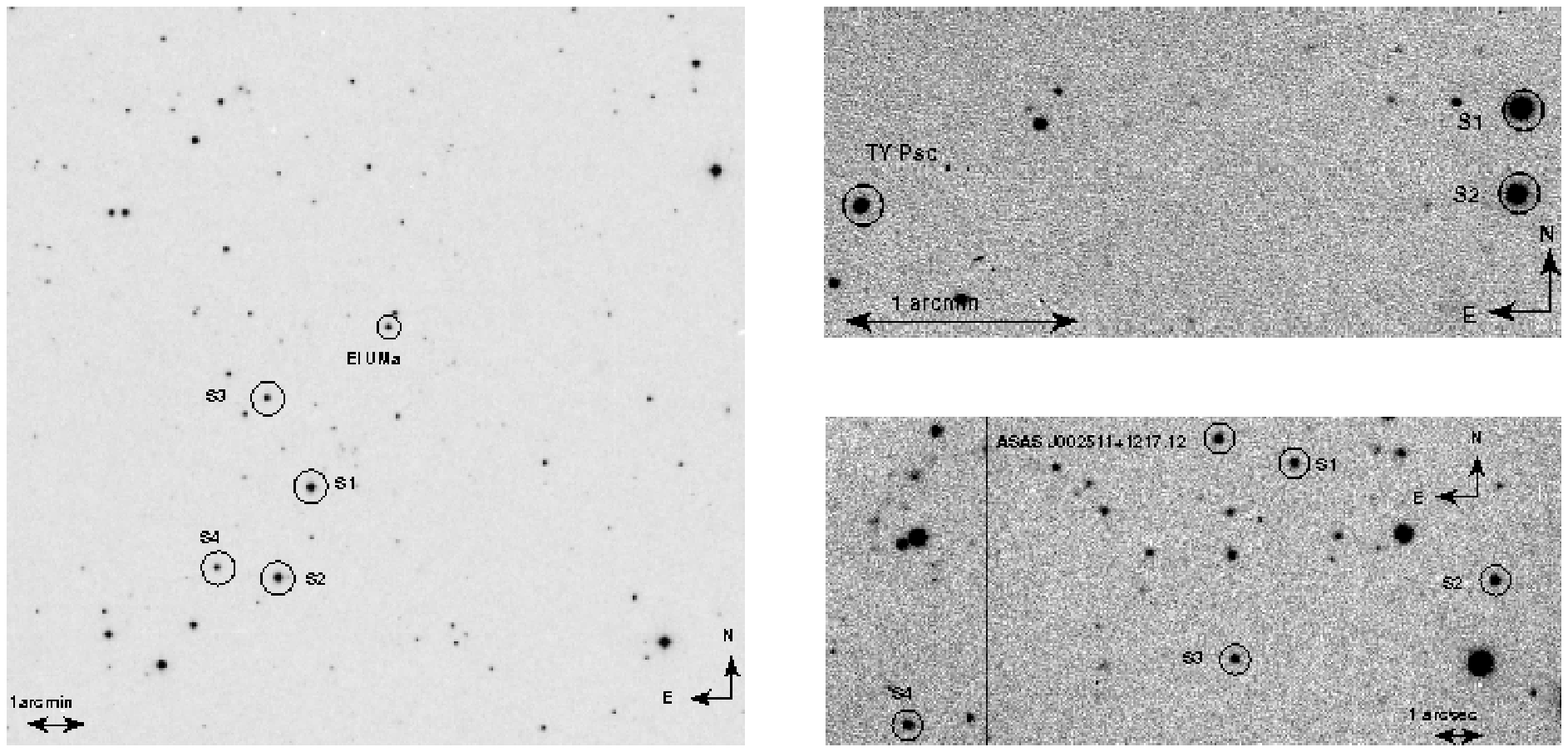}}
      \caption{ ({\sl continued}) The finding charts of EI UMa, TY Psc and ASAS J002511+1217.12. Marked on each chart are the comparison stars that served for differential photometry.}
      \label{f_fov_chart2}
    \end{figure}

\section{Data reduction}          

The CCD frames were processed for bias removal and flat-field correction. Small sets of bias frames were taken at regular time intervals during the night, and were then combined into one median nightly bias frame, which was then subtracted from  all images. In case of significant drops in the CCD temperature, new bias frames were obtained at that point and neighbouring images were processed with the new combined bias frame. For each night we obtained a median flat-frame. In case of bad weather during both evening and morning twilight, the flat frame of the closest observing night was used.
    
Aperture photometry was made using the IRAF package {\tt apphot}. The circular aperture radius used for  the computation of the instrumental magnitudes was equal to $2\times{\it FWHM}$. When the nights were photometric and stable, a mean  ${\it FWHM}$ was calculated, resulting in a fixed nightly aperture. In case of unstable sky quality with strongly variable  ${\it FWHM}$,  the night was divided into parts and each part was treated independently by calculating the corresponding aperture. After extracting the instrumental magnitudes of the CV and the selected comparison stars, the following procedure was applied in order to check the constancy of the comparison stars. The behaviour of the  instrumental differential magnitudes of the comparison stars throughout the campaign was checked and compared to the difference of the instrumental magnitudes between the CV and one comparison star. This was done by plotting the differential magnitudes against time. Only when the series of differential magnitudes of a set of  comparison stars  did not reveal any night to night trend, and when  the standard deviation $\sigma$ of the comparison stars' differential magnitudes  was at least 3 times smaller than that of the CV's were the corresponding comparison stars retained for further use. Differential photometry was then applied according to the following procedure. We first calculated the average instrumental magnitude of each comparison star along with its $\sigma$ and then determined  the weighted average of the instrumental magnitudes of all comparison stars. The weights for each comparison star were set equal to the inverse  of their $\sigma$, and  the differential magnitudes of each CV  were then obtained by subtracting the weighted average  magnitude from the corresponding instrumental magnitude of the CV. 

Given the fact that no filters were used, or if used had to be compared to unfiltered data, it was not possible to obtain proper  transformations to a photometric system. However, the following procedure, that yields the most reliable magnitudes in our case and provides the possibility of comparing to future or past data, was applied. From the comparison stars we chose those for which \citet{hen95} give $V$ magnitudes. If their catalogue did not contain the required stars, the respective magnitudes were taken from the online USNO-A2.0 (United States Naval Observatory) catalogue at ESO (European Southern Observatory) science archive facility. For each observing night the catalogued  magnitude of each comparison star  was plotted against the differential magnitude (resulting from the mean of its differential magnitudes throughout the night) and through linear least-squares fitting an equation that best fitted them was extracted. If all nights belonged to the same observing run and site then a mean equation was calculated and therefore the differential magnitudes of our CV were transformed to magnitudes that are approximately compatible with the magnitude scale of the catalogue. When runs at different observing sites were involved, the behaviour of the magnitude difference between two comparison stars (common in the FOV of all runs) was checked and the appropriate magnitude shift was applied. In this way we were able to compare all light curves obtained at different observatories. 
    
The same reduction technique and light curve generation was applied to all CVs examined in this study. Table~\ref{t_comps} gives the designations of the comparison stars used for each target, as seen in Fig.~\ref{f_fov_chart1}, along with their magnitudes and source from the literature. Whenever any offset occurs it will be stated in the corresponding section.

For comparison reasons and in order to have an impression of how real each system's nightly mean-brightness variations are, Table~\ref{t_behaviour} gives $\mu_{\rm CV}$, which is each system's mean nightly magnitude, and $\mu_{\rm dif}$. The mean nightly differential magnitude between two comparison stars common in the FOV of all observing runs. The comparison stars, from which  $\mu_{\rm dif}$ is calculated for each case, are also given. 

In addition, when plotting the light curves of all objects in the next sections, an error bar is shown in the box of each night's light curve. This serves as an estimation of the error on a single measurement of the magnitude of the CV and is represented by the standard deviation around the mean of the differential magnitudes,  $\sigma_{(m_{S_i}-m_{S_j})}$ between two comparison stars $S_i$, $S_j$. However, this can only be true if the appropriate $S_i$, $S_j$ are chosen. The ideal $S_i$, $S_j$ are those that a similar magnitude as that of the CV (there are cases where the magnitudes of the comparison stars differ by 0.5--2\,mag from that of the CV). The use of   comparison stars  that are that much fainter would lead   to either overestimation or underestimation of the  error. So, for each night the following procedure was adopted in order to apply the appropriate correction. We first calculated the mean and corresponding $\sigma$ of the quantity $m_{S_i}-m_{S_j}$, where $m_{S_i}$, $m_{S_j}$ are the magnitudes of $S_i$ and $S_j$. Taking into account the magnitude of the CV, $m_{S_i}$, $m_{S_j}$ and Table~\ref{t_comps}, we found another night during which four comparison stars, two ($S_k$, $S_l$) with magnitudes similar to $m_{S_i}$ and $m_{S_j}$, and two ($S_m$, $S_n$) with magnitudes similar to that of our CV, co-existed. We then calculated the correction factor:

\begin{equation}
A=\frac{\sigma_{(m_{S_m}-m_{S_n})}-\sigma_{(m_{S_k}-m_{S_l})}}{\sigma_{(m_{S_k}-m_{S_l})}}
\end{equation}

This correction factor now gives us the opporunity to pass from the available $\sigma_{(m_{S_i}-m_{S_j})}$ to $\sigma_{(m_{S_i}-m_{S_j})}^{'}$ that we would have, had $m_{S_i}$, $m_{S_j}$ been close to the magnitude of the CV. The link between them is given by the equation:

\begin{equation}
\sigma_{(m_{S_i}-m_{S_j})}^{'}=\sigma_{(m_{S_i}-m_{S_j})}(1+A)
\end{equation}

This quantity now well matches the error on a single measurement of each CV and is represented by the error bar in each light curve. Moreover, what should be pointed out is that due to the change of the FOV between different observing runs of the same CV, $S_{i,j,k,l,m,n}$ are not necessarily the same between all runs of the same target. 

As will be easily noticed from the error bars of the light curves in the upcoming sections, in most of the cases there are small variations between the nights. These variations are expected due to variations of the weather and therefore atmospheric conditions. Furthermore, if one CV appears in quiescence during one observing run and in outburst during another, the error is expected to decrease supposing there are not any great differences in the atmospheric conditions. However, the observing site of UC is always   subjected to greater errors than the other sites because  UC is located in a suburb  of Brussels and therefore   operates under much poorer atmospheric conditions compared to the rest of the observing sites we used. In addition, its low altitude makes it even more vulnerable to the variations of the lower atmosphere. These are the reasons of the large variance in the errors shown for some CVs. For V844 Her and LN UMa, the error estimation at UC is 10 and 4 times, respectively, greater than that at SK. For EI UMa, which we observed simultaneously and under good weather conditions at UC and HL, we found that the error at UC is 3.5 times greater than that at HL. V516 Cyg, observed only at UC, also shows an increase in the error but this is attributed not only to poorer atmospheric conditions but most predominantantly to its decrease in brightness by 2\,mag.

\begin{table}[!t]
\caption{Designation of comparison stars.}
\label{t_comps}
\centering
\begin{tabular}{|ll|ll|}
\hline
 {\it EI UMa}	 &		 &  {\it GZ Cnc}   &		  \\
S1: [HH95] EI UMa-10& $V$=13.406 &  S1: U0975\_06195484& $R$=15.7 \\
S2: [HH95] EI UMa-4 & $V$=13.403 &  S2: U0975\_06196461& $R$=15.5 \\
S3: [HH95] EI UMa-23& $V$=15.922 &  S3: U0975\_06197019& $R$=14.0 \\
S4: [HH95] EI UMa-32& $V$=15.168 &  S4: U0975\_06197135& $R$=13.0 \\
		 &		 &  S5: U0975\_06195399& $R$=14.7 \\
{\it V844 Her}     &		 &  S6: U0975\_06194753& $R$=15.1 \\
S1: U1275\_08930994& $R$=13.5	 &  S7: U0975\_06195450& $R$=14.9 \\
S2: U1275\_08930976& $R$=15.7	 &  S8: U0975\_06195694& $R$=15.9 \\
S3: U1275\_08931020& $R$=15.4	 &  S9: U0975\_06196184& $R$=15.4 \\
S4: U1275\_08931359& $R$=14.4	 &  S10: U0975\_06195886& $R$=15.3\\
S5: U1275\_08932829& $R$=14.7	 &                      &           \\
		 &		 &  {\it TY Psc}& \\ 		      
{\it V751 Cyg} &                 &  S1: [HH95] TYPsc-24 &$V$=15.240\\     		    
S1: [HH95] V751Cyg-7& $V$=13.552 &  S2: [HH95] TYPsc-23 &$V$=15.925\\
S2: [HH95] V751Cyg-3& $V$=12.284 &  		   &	      \\
S3: U1275\_14247362& $R$=14.3    &  {\it ASAS J002511+1217.12} &\\
S4: U1275\_14251141& $R$=12.2    &  S1: U0975\_00086988 &$R$=15.9\\
S5: U1275\_14248730& $R$=12.3    &  S2: U0975\_00086066 &$R$=15.4\\
S6: U1275\_14247527& $R$=13.7    &  S3: U0975\_00087221& $R$=16.0\\
S7: U1275\_14260111& $R$=16.3    &  S4: U0975\_00088736& $R$=16.1\\	       
	           &             &   	     &         \\
{\it V516 Cyg}	&                   &  {\it V1315 Aql}& \\
S1: U1275\_14143883& $R$=13.9       &  S1: [HH95] V1315Aql-23& $V$=14.895\\
S2: U1275\_14144823& $R$=14.7       &  S2: [HH95] V1315Aql-38& $V$=15.992\\
S3: U1275\_14145851& $R$=14.8       &  S3: U0975\_14381457& $R$=16.3\\       
S4: U1275\_14142515& $R$=13.6	    &			     &         \\
          &                         &  {\it LN UMa}& \\
 	  &		 	    &  S1: [HH95] PG1000+667-13& $V$=14.307\\
   	  &			    &  S2: [HH95] PG1000+667-22& $V$=15.527\\
   	  &			    &  S3: [HH95] PG1000+667-21& $V$=15.859\\
   	  &			    &  S4: [HH95] PG1000+667-19& $V$=16.209\\				 	  
  \hline			 
 \end{tabular}
\begin{center}
{\footnotesize 
	    Notes: Designations starting with HH95 refer to stars as taken from \citet{hen95}. \\The rest are adopted from the USNO-A2 catalogue.\hfill}
\end{center}
\end{table}

\begin{table}
\caption{Long-term behaviour.}
\label{t_behaviour}
\centering
\begin{tabular}{|lll|lll|}
\hline
UT date & $\mu_{\rm CV} \pm \sigma_{\rm CV}$ & $\mu_{\rm diff} \pm \sigma_{\rm diff}$ & UT date & $\mu_{\rm CV} \pm \sigma_{\rm CV}$ & $\mu_{\rm diff} \pm \sigma_{\rm diff}$ \\
\hline
{\it EI UMa OHL}&	   &  $\mu_{(\rm S4-\rm S1)}$ &  {\it GZ Cnc}    & 	  & $\mu_{(\rm S1-\rm S2)}$\\
09dec02     &  15.096$\pm$0.054  & 1.694$\pm$0.012		   &  09mar03	      & 15.415$\pm$0.140   & 0.193$\pm$0.046 \\		
10dec02     &  14.964$\pm$0.079  & 1.688$\pm$0.013		   &  16mar03	      & 13.911$\pm$0.070   & 0.205$\pm$0.143	\\	
31jan03     &  14.564$\pm$0.044  & 1.721$\pm$0.018		   &  22mar03	      & 15.223$\pm$0.044   & 0.231$\pm$0.054	\\	
	    &			 &		  		   &  25mar03	      & 14.545$\pm$0.098  & 0.221$\pm$0.123	\\	
{\it EI UMa UC}&	  &   $\mu_{(\rm S3-\rm S1)}$ &                  &                        &       \\
31jan03     &  14.573$\pm$0.050  & 2.510$\pm$0.044		   & {\it TY Psc}&	&  $\mu_{(\rm S2-\rm S1)}$ \\
14feb03     &  15.057$\pm$0.039  & 2.238$\pm$0.031		   & 21jul03	 &  17.695$\pm$0.076  & 0.612$\pm$0.004 \\ 
16feb03     &  14.431$\pm$0.037  & 2.232$\pm$0.034		   & 22jul03	 &  17.514$\pm$0.106 &  0.599$\pm$0.008 \\  
17feb03     &  14.451$\pm$0.046  & 2.239$\pm$0.023	    	   & 23jul03	 &  17.597$\pm$0.100  & 0.604$\pm$0.003  \\
	    &			 &		  		   & 10jan05	 &  14.066$\pm$0.051  & 0.518$\pm$0.013 \\ 
{\it V844 Her} &	  &   $\mu_{(\rm S4-\rm S1)}$ &		 &		      & 		\\ 
04jul01     &  16.403$\pm$0.185  & 0.879$\pm$0.019		   & {\it J002511}      &  	   &			\\		 
15may02     &  16.454$\pm$0.102  & 0.881$\pm$0.012		   & {\it +1217.12} &	   & $\mu_{(\rm S1-\rm S2)}$ \\
16may02     &  16.420$\pm$0.105  & 0.882$\pm$0.011		   & 07oct04	 &  15.604$\pm$0.062  & 0.392$\pm$0.025 \\		 
12jun02     &  16.136$\pm$0.046  & 0.897$\pm$0.003	    	   & 10oct04	 &  15.672$\pm$0.133  & 0.342$\pm$0.033  \\		 
14jun02     &  16.097$\pm$0.082  & 0.872$\pm$0.009		   & 11oct04	 &  15.761$\pm$0.066  & 0.355$\pm$0.021 \\		 
15jun02     &  16.204$\pm$0.044  & 0.869$\pm$0.007		   &		 &		      & 		 \\
	    &			 &		  		   & {\it V1315 Aql} &  	   & $\mu_{(\rm S2-\rm S1)}$\\
{\it V751 Cyg}&	          &  $\mu_{(\rm S7-\rm S1)}$                 & 14aug03	     & 15.134$\pm$0.075  & 0.560$\pm$0.004\\	 
14may04     & 14.291$\pm$0.051   & 1.572$\pm$0.019		   & 15aug03	     & 14.552$\pm$0.082  & 0.568$\pm$0.008 \\		 
15may04     & 14.401$\pm$0.060   & 1.541$\pm$0.024		   & 16aug03	     & 15.111$\pm$0.084  & 0.567$\pm$0.004  \\		 
16may04     & 14.458$\pm$0.050   & 1.536$\pm$0.022		   & 17aug03	     & 14.987$\pm$0.093  & 0.562$\pm$0.004 \\		 
24may04     & 14.324$\pm$0.053   & 1.534$\pm$0.038		   & 31aug04	     & 14.839$\pm$0.096  & 0.613$\pm$0.007 \\	 
26may04	    & 14.320$\pm$0.059   & 1.544$\pm$0.024		   & 01sep04	     & 14.882$\pm$0.124  & 0.612$\pm$0.007 \\		 
27may04     & 14.403$\pm$0.052   & 1.551$\pm$0.023		   & 02sep04	     & 14.796$\pm$0.161  & 0.616$\pm$0.006 \\		 
28may04     & 14.356$\pm$0.051   & 1.545$\pm$0.023		   &		     &  		&		   \\
29may04     & 14.304$\pm$0.066   & 1.547$\pm$0.038		   & {\it LN UMa}      &	     & $\mu_{(\rm S3-\rm S2)}$ \\
07may05     & 14.606$\pm$0.069   & 1.641$\pm$0.018		   & 18feb03	     & 15.308$\pm$0.055  & 0.242$\pm$0.040 \\	
08may05     & 14.427$\pm$0.088   & 1.633$\pm$0.015		   & 25feb03	     & 15.207$\pm$0.056  & 0.253$\pm$0.033 \\	
09may05     & 14.318$\pm$0.058   & 1.634$\pm$0.009                 & 01jun04	     & 15.401$\pm$0.018  & 0.264$\pm$0.008 \\	
13may05     & 14.407$\pm$0.065   & 1.624$\pm$0.015		   & 02jun04	     & 15.343$\pm$0.021  & 0.261$\pm$0.012 \\	
            &                    &                	    	   & 06jun04	     & 15.293$\pm$0.026  & 0.265$\pm$0.006 \\	
{\it V516 Cyg}  &          &$\mu_{(\rm S2-\rm S1)}$   &              &                   &                 \\
08oct02         & 13.585$\pm$0.013  & 1.008$\pm$0.024		   &		      & 		  &		    \\
09oct02         & 13.609$\pm$0.013  & 1.012$\pm$0.022		   &		      & 		  &		    \\
19oct02         & 15.304$\pm$0.064  & 1.004$\pm$0.019		   &                 &                   &                 \\	       
23oct02         & 15.806$\pm$0.040  & 1.007$\pm$0.013		   &                 &                   &                 \\ 	       
05nov02         & 15.885$\pm$0.078  & 0.987$\pm$0.022		   &                 &                   &                 \\  	                    	     
  \hline
  \end{tabular}
 \begin{center}
 {\footnotesize 
 	     Notes: $\mu_{\rm CV}$ and $\sigma_{\rm CV}$ are the mean nightly magnitude and corresponding $\sigma$ of each CV  ; $\mu_{\rm dif}$ and $\sigma_{\rm dif}$ are the mean nightly differential magnitude and corresponding $\sigma$ between two compariosn stars. At the top of each $\mu_{\rm dif}$ column, $\mu_{\rm dif}$ is explicitly written, indicating which comparison stars were used in each case.\hfill}
 \end{center}
 \end{table}

\section{EI UMa}
      EI UMa (PG 0834+488) was discovered in the Palomar-Green survey \citep{gre82} and was classified as a CV on the basis of its optical spectrum which showed Balmer and high-excitation emission lines. Emission from high-excitation lines is characteristic of magnetic CVs. \citet{coo85}, in the first X-ray study on this target, found that EI UMa is a hard X-ray source with a low column-density and an X-ray-to-optical density resembling that of DN CVs. This, combined with the facts that the value of the column density matched that of non-magnetic CVs and, most importantly, that there was no X-ray modulation, drew him to the conclusion that EI UMa better matches a DN. One year later, \citet{tho86} performed the first radial-velocity study and detected a 6.4-h orbital period ($P_{\rm orb}$), the spectrum resembling that of a DN. On the other hand, he was of the opinion  that the rather strong HeII $\lambda$4686 emission, the X-ray characteristics and the system magnitude, implied a possible DQ Her type. One night of photoelectric photometry \citep{wil86} showed small amplitude flickering but no larger-amplitude long-term modulations.

\begin{figure}[!ht]
  \centerline{\includegraphics[width=1\textwidth]{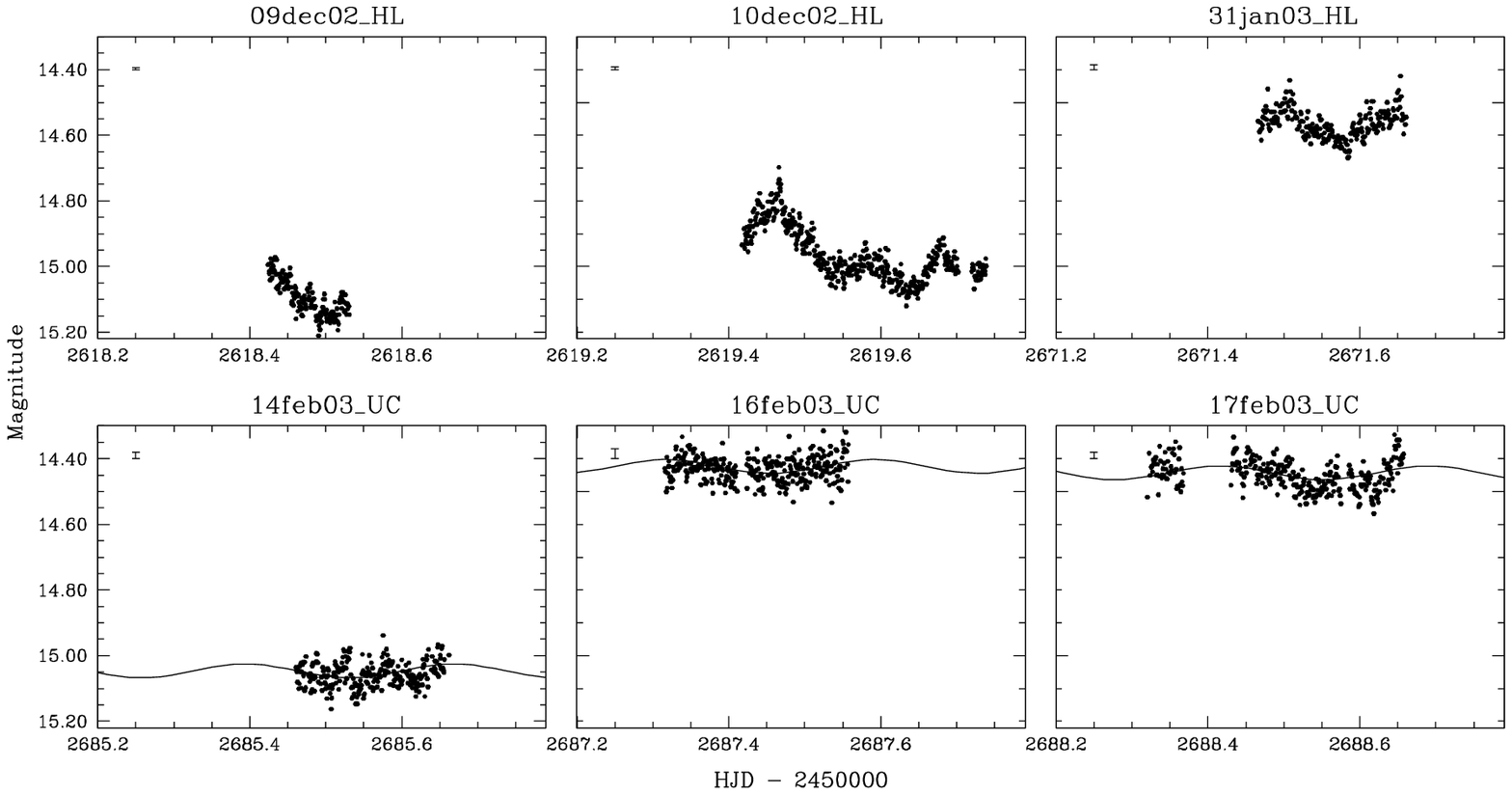}}
      \caption{EI UMa light curves. In the upper left part of each light curve the error on a single measurement is indicated. The sine curve with $  P=3.598$\,\rm c \rm d$^{-1}$ is superimposed on the UC light curves.}
      \label{f_lc_EIUMa}
    \end{figure}

Our dataset consists of observations obtained at OHL (3 nights) and at UC (3 nights). All observations were unfiltered and the exposure time varied from 45 to 90\,sec. Due to the lack of a common couple of comparison stars at the two sites, the overlapping observing night of 31 January 2003 was used for the comparison between the OHL and UC light curves. This was done by equating the mean magnitude of our system in the overlapping period of time at the two runs. The light curves are shown in Fig.~\ref{f_lc_EIUMa}. The OHL and UC runs in Table~\ref{t_behaviour} are separated since $\mu_{\rm diff}$ does not correspond to the same couple of comparison stars. 

In Fig.~\ref{f_lc_EIUMa} it is evident that the system's brightness, in both runs, showed an increase by $\approx$0.6\,mag. A closer look at Table~\ref{t_behaviour} points to the reality of this increase since the variation of the mean differential magnitude between the comparison stars is of a much smaller scale. However, we can see that between the two runs a great change in the system's behaviour occurs. For the subsequent analysis we have therefore divided our dataset into 2 subsets, one for each observing run. 

For EI UMa, as well as the rest of the CVs that are presented in this study, frequency analysis was carried out by using the {\tt period04} package \citep{len05}, based on the Discrete Fourier Transform method. The calculation of the uncertainties of the corresponding parameters was performed by means of Monte Carlo simulations from the same package. Before performing the frequency analysis, the trend in the light curves was removed by bringing them around their nightly mean. For the first dataset we find a periodicity of 2.7911$\pm$0.0002\,$\rm c \rm d^{-1}$ with a semi-amplitude of 0.076$\pm$0.002\,mag, along with the second highest peak, its 1-d alias. However, the 1-d alias of the periodicity we find is the spectroscopic $P_{\rm orb}$. This suggests that even though our data are better fitted by a periodicity of 2.7911$\pm$0.0002\,$\rm c \rm d^{-1}$, its 1-d alias (the spectroscopic period) is more likely the right period. When running frequency analysis on the residuals, a periodicity of 9.57\,$\rm c \rm d^{-1}$ or 2.5\,h shows up, obvious in our light curves. However, this signal does not occur on a stable time-scale and its folding is not promising. Given the short duration of the light curves we cannot draw conclusions on its origin, since it could just be a short-lived modulation as seen in many CVs. 

The second dataset has more to reveal. The 2.5-h variability seen in the first dataset is absent, therefore making it easier to detect the $P_{\rm orb}$. Indeed, we have detected a modulation at 3.640$\pm$0.011$\rm c \rm d^{-1}$ or 6.59\,h which agrees with the spectroscopic $P_{\rm orb}$ of 6.43\,h. Its low semi-amplitude of 0.021$\pm$0.002\,mag, clearly shows the reason why it could not be detected in the first subset. Moreover, it is true that had the periodicity not been detected spectroscopically, it would have been difficult to trust it. The sinusoidal representation of the periodicity has been superimposed on the UC light curves of Fig.~\ref{f_lc_EIUMa}.
 
\begin{figure}[!ht]
  \centerline{\includegraphics[width=1\textwidth]{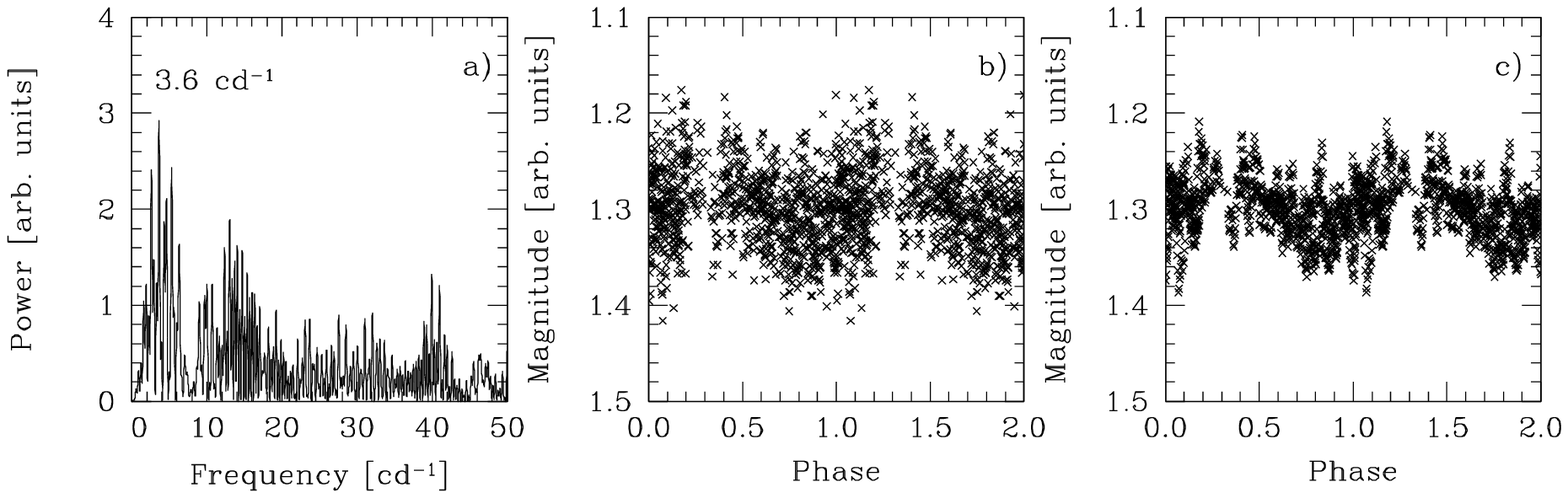}}
      \caption{a) Resulting power spectra of the second subset (UC run) with the dominant frequency indicated b) Data folded on the 6.67-h periodicity c) Folded data, smoothed with a boxcar of 5. }
      \label{f_ps_EIUMa}
    \end{figure}

The amount of flickering was also found from the standard deviation $\sigma$ in each light curve. Then the average of all $\sigma$ and its  $\sigma$ as uncertainty, were computed for both subsets. For the OHL run it was found equal to 0.059$\pm$0.018 and for the UC run 0.043$\pm$0.006. The amount of flickering decreased with the increasing brightness between the two runs, a behaviour also noticed between the individual nights. 

We also searched for possible Quasi Periodic Oscillations (QPOs). These signals, unstable in both amplitude and frequency can be omitted by Fourier analysis. We therefore apply the following procedure. We average the power spectra of all nights, excluding very short ones if present, and then plot them in log-log scale. In this way, possible QPOs, will reveal themselves as broad features. In the case of EI UMa this was done for both subsets and the whole set as well, however the variations seemed characteristic of flickering and no QPOs were evident. The fact that a small number of nights is averaged is not in favour of a QPO detection.

\section{V844 Her}

V844 Her is a DN discovered by \citet{ant96} on photographic plates. The
best observed outburst lasted between 12 and 18 days. Since then, various variable
star organisations followed this object. During its first outburst, clear
superhumps were visible  \citep{sco96,van96} and during a superoutburst a period
of 0.056$\pm$0.001\,d was found. More precise measurements of the superhump
period ($P_{\rm sh}$) are given by \citet{kat00} and \citet{tho02} during the 1999 and 1997 outbursts, respectively. As \citet{tho02} noted, the eruption timescale, a maximum brightness plateau of more than 10\,d with a following rapid decrease in 1--2\,d, as well as the superhumps establish the star's membership in the SU UMa class of DN. Quiescent spectra, part of the same study, appeared typical of those of DN and had double-peaked emission lines. The $\rm P_{\rm orb}$, measured from  the H$\alpha$ emission-line velocities, was found to be 0.054643$\pm$7$\cdot 10^{-6}$\,d.

\begin{figure}[!ht]
  \centerline{\includegraphics[width=1\textwidth]{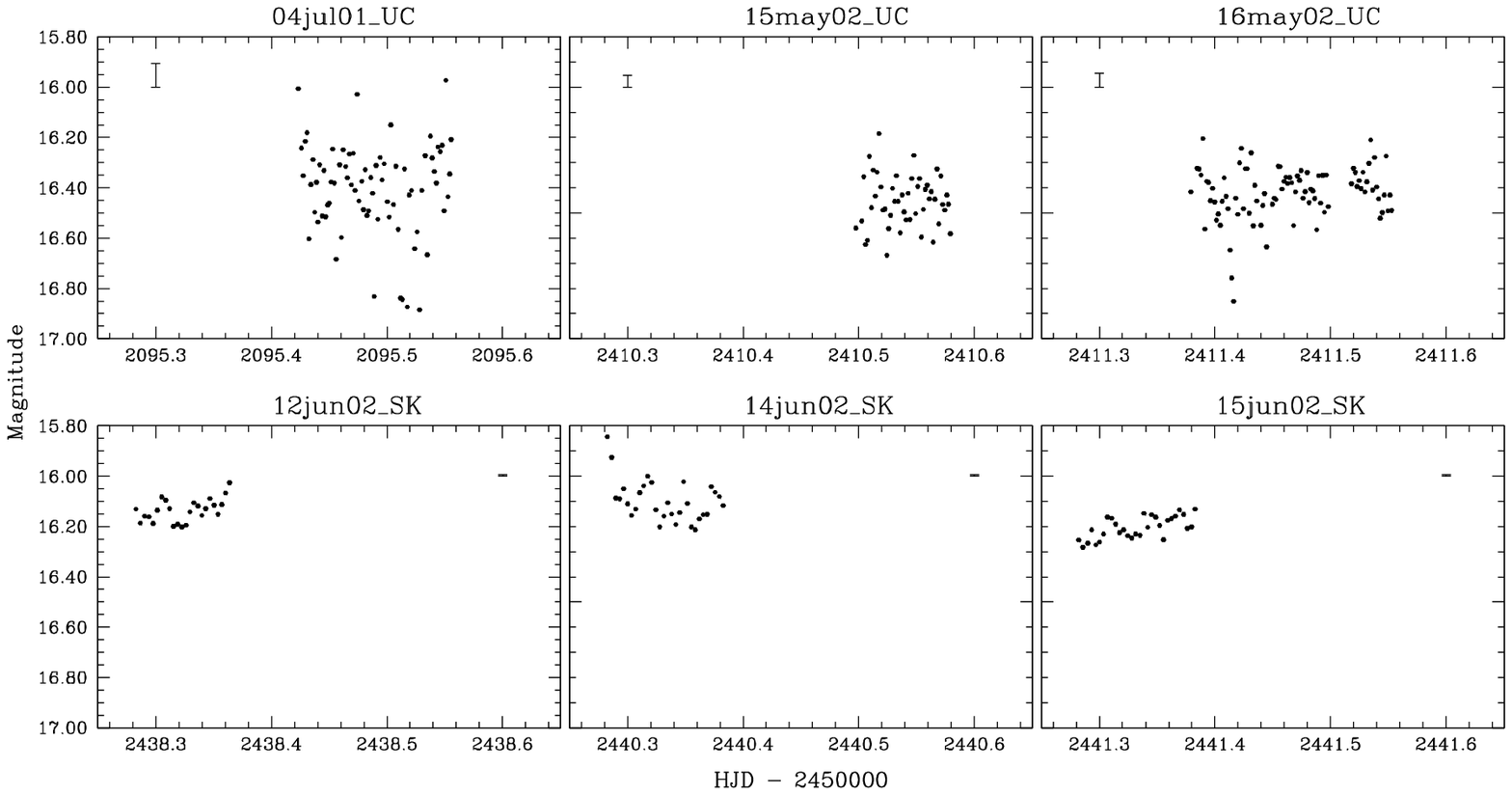}}
      \caption{V844 Her light curves. In the upper left (or right) corner of each light curve the error on a single measurement is indicated.}
      \label{f_lc_V844Her}
    \end{figure}

V844 Her was observed during 3 nights at UC and 3 at SK, all unfiltered. The exposure time varied between 120 and 240\,sec. The light curves are shown in Fig.~\ref{f_lc_V844Her}. The error bars on the upper left (or right) corners of each light curve indicate the error on a sinlge measurement. 

We performed frequency analysis, but no significant periodicity was detected. The amount of flickering activity was found to be 0.131$\pm$0.047\,mag for the UC run and 0.057$\pm$ 0.022\,mag for the SK one. However, the mean $\sigma$ of the comparison stars, as indicated in the upper left part of each light curve in Fig.~\ref{f_lc_V844Her}, is greater in the UC run by a similar factor as the difference in the amount of flickering. Therefore, we do not think that the flickering between the two runs has significantly changed and only the amount found for the SK should be trusted. 

An inspection of the AAVSO long-term light curves \footnotemark[1] \footnotetext[1]{The AAVSO light curve generator tool can be found at http://www.aavso.org/data/lcg/. Using as input our object, JD 2452400--2452450 and by marking only the visual-validated data to appear, the superoutburst light curve is generated. Another superoutburst light curve but with a more complete coverage can be found for JD 2452150--2452181.} revealed that we have just missed a superoutburst in the meantime beween the two observing runs. The nights of 15 and 16 May 2002 are placed just before the initiation or at the rising part of the superoutburst. During the nights of 12--15 June 2002 the system has returned to quiescence. 

\section{V751 Cyg}
 V751 Cyg (EM* LkH$\alpha$) lies near the ``eye'' of the Pelican nebula (IC 5070). Its variability  was discovered by \citet{mar58} and initially believed to be an R Coronae Borealis star. Later spectra \citep{her58,her72} revealed its CV nature. This system had already shown drops up to 2.5\,mag, but it was only after high-speed photometry was obtained, that it was first suggested to be a NL of the VY Scl subtype \citep{rob74}. This suggestion was later confirmed by \citet{dow95} and \citet{mun97}, on the basis of spectra showing Balmer and He emission (or emission cores in some cases), albeit not of great strength except H$\alpha$. V751 Cyg was also identified, during a low state, as a transient supersoft X-ray source \citep{gre99}. X-ray observations during the high state showed that this was not the case. An anti correlation of X-ray and optical intensity was also observed. Optical spectra during the same low-state revealed a very blue continuum and strong emission lines (Balmer, HeI and HeII $\lambda$4686).
 
The first detailed spectrophotometric study was published by \citet{pat01}. They reported photometric signals at 3.348\,h or $f_1=7.170\pm0.003\,\rm c\rm d^{-1}$and 3.9\,d or $f_2=0.254\pm0.005\,\rm c \rm d^{-1}$ as well as a modulation of the radial velocities at 3.468\,h. These signals were identified as a negative superhump period $P_{\rm sh}^-$, the wobbling period of a tilted accretion disc $P_{\rm wob}$ and the $P_{\rm orb}$, respectively. Additionally a possible QPO around 20\,min was reported. Occasionally the system showed P Cygni absorption profiles in the Balmer and HeI lines, correlated to the binary phase.

\begin{sidewaysfigure}[p]
 \centerline{\epsfig{figure=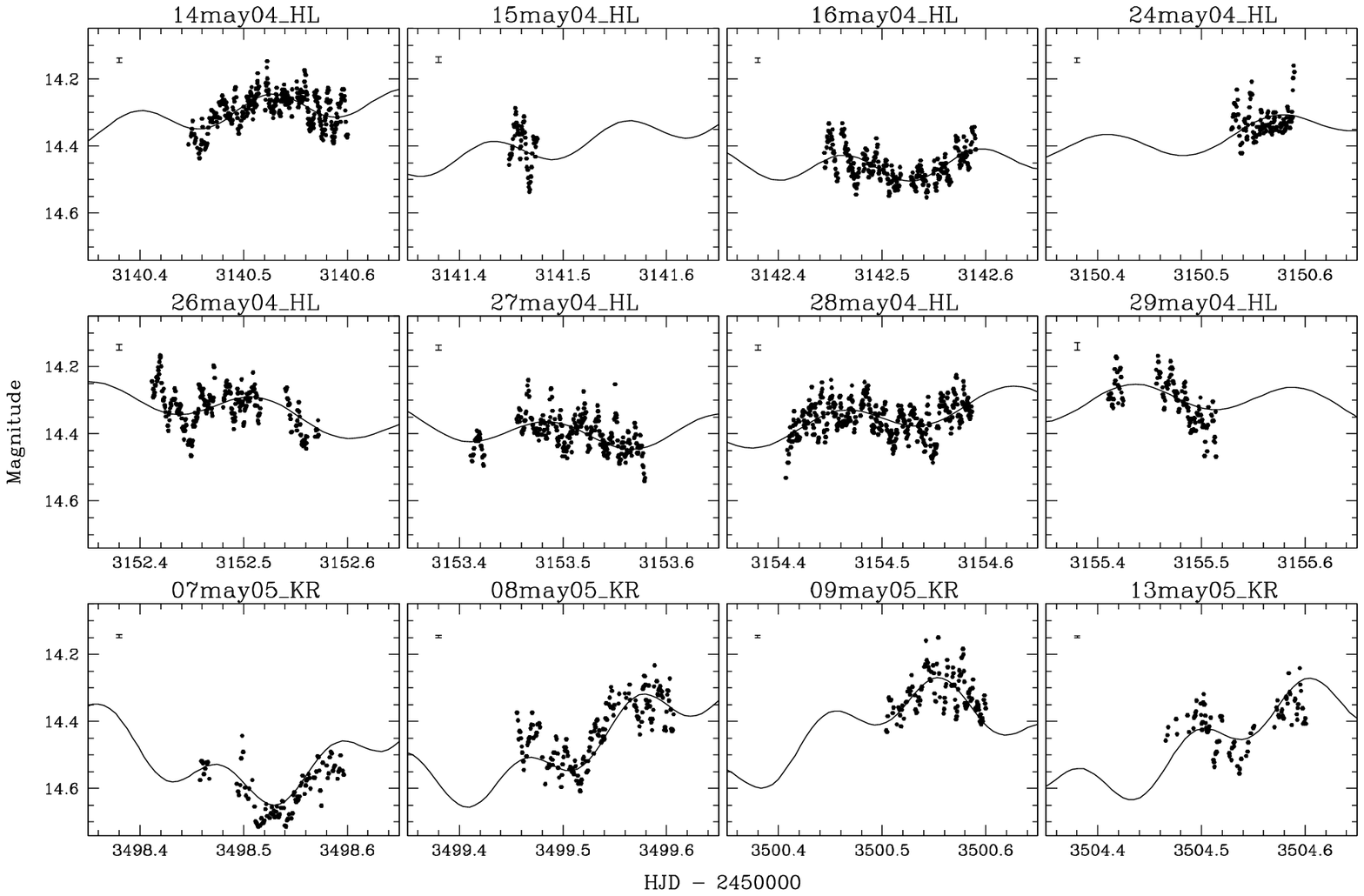,width=1\textheight,height=.65\textwidth}} 
 \caption{V751 Cyg light curves with the corresponding errors idicated in the upper left corner of each light curve. The sine curves, superimposed on the light curves, represent the periodicities found in each subset.}
 \label{f_lc_V751Cyg}
 \end{sidewaysfigure}

Our dataset consists of 8 nights of observing at OHL and 4 nights at KR. All runs were unfiltered and the exposure time varied between 20 and 80\,sec, depending on the telescope and atmospheric conditions. The light curves are shown in Fig.~\ref{f_lc_V751Cyg}.

\begin{figure}[!ht]
  \centerline{\includegraphics[width=1\textwidth]{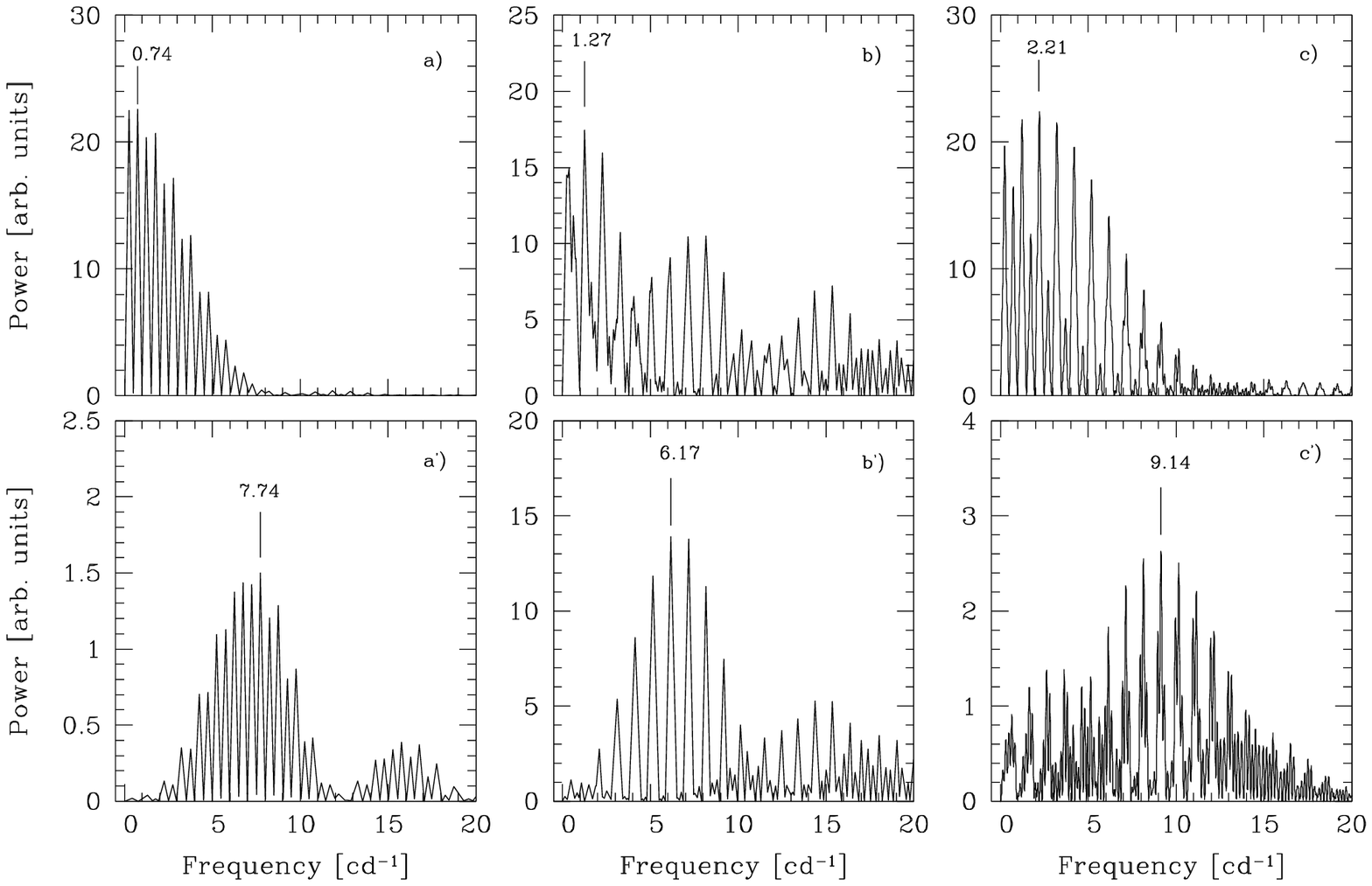}}
      \caption{V751 Cyg power spectra with the dominant frequencies indicated. a, b and c correspond to the power spectra of the first, second and third subset, respectively. a$^\prime$,  b$^\prime$ and c$^\prime$ are the power spectra of the residuals of the first, second and third subset, respectively. 
}
     \label{f_ps_V751Cyg}
    \end{figure}

The photometric frequencies, found by \citet{pat01}, are also detected in our data, albeit as their aliases or harmonics in some cases. Before applying frequency analysis, our dataset was divided into 3 subsets. The first includes the 3 first nights of the OHL run, the second the next 5 nights and the third the KR run, one year later than the OHL one. In all cases we cannot directly detect the low frequency $f_2$. In the first subset we find its $3^{\rm rd}$ harmonic, in the second its 1-d alias and in the third its 2-d alias. In contrast to \citet{pat01}, the number of each subset's consecutive observing nights is not larger than 4. This, makes the direct detection of the $\approx$4-d $P_{\rm wob}$ more difficult. Performing period analysis on the residuals revealed the existence of the $P_{\rm sh}^-$. The corresponding frequencies for the first, second and third subset are 7.739$\pm$0.005, 6.171$\pm$0.002 and 9.144$\pm$0.004\,$\rm c \rm d^{-1}$, respectively. Those signals reflect the combination of frequencies $f_1+2 f_2$ (or an harmonic of the first frequency found in the same subset), the 1-d alias of $f_1$, and the 2-d alias of  $f_1$, respectively. The power spectra of the original and residual data are shown in the top and bottom row of Fig.~\ref{f_ps_V751Cyg}. Each column from left to right corresponds to the first, second and third subset, respectively. Both periodicities found in each subset are superimposed on the lights curves of Fig.~\ref{f_lc_V751Cyg}. The aliases of both periodicities as well as the harmonics of the $\rm P_{\rm sh}^{-}$ are evident. As seen in Fig.~\ref{f_ps_V751Cyg}a--Fig.~\ref{f_ps_V751Cyg}c, the 4-d aliases are so strong that dominate the periodogram in the vicinity of $P_{\rm sh}^-$ and bias its estimate. We therefore first fitted the 4-d period, then subtracted it and finally ran periodogram analysis again. Once more the aliases appeared stronger. We also applied the same procedure in the whole of the OHL run (i.e. subset 1 and 2) but again we could not detect $f_1$ but its 1-d alias. Treating all light curves as a whole, once more reveals the 1-d alias of $f_2$ and the 1-d alias of the $\rm P_{\rm sh}^{-}$ when period analysis is run on the residuals. After subtracting all periodic signals the amount of flickering was measured and found to be 0.048$\pm$0.007\,mag.

\begin{figure}[!ht]
  \centerline{\includegraphics[width=1\textwidth]{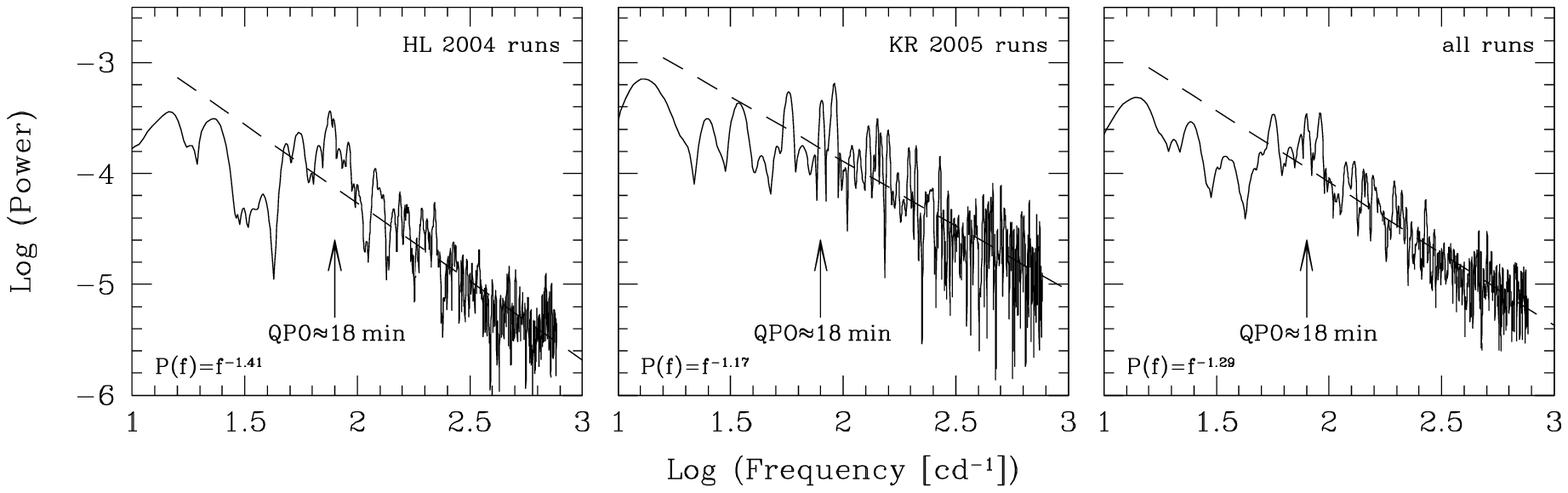}}
      \caption{From left to right: V751 Cyg average power spectra for the OHL 2004 run, KR 2005 runs and all runs. The fit of the linear parts of the power spectra along with the corresponding equations are shown in the lower left part of each graph. }
      \label{f_qpo_V751Cyg}
    \end{figure}

A closer look at Fig.~\ref{f_lc_V751Cyg} reveals that there must be high frequency oscillations. A QPO at 20\,min has already been reported by  \citet{pat01}. We therefore applied the same procedure, as in the case of EI UMa, to search for oscillations. Since the OHL and KR runs are one year apart, and from the light curves of Fig.~\ref{f_lc_V751Cyg} we notice a different behaviour of the high frequency variations, we have treated them separately. Fig.~\ref{f_qpo_V751Cyg} shows the average power spectra in each run as well as all runs together. It is now clear that the two runs differ. In the OHL run a broad figure is evident from 63\,$\rm c\rm d^{-1}$ on, with a peak around 80\,$\rm c\rm d^{-1}$. This $\approx$18\,min QPO is the counterpart of the high frequency variations seen in the OHL light curves. In contrast to the OHL run, the KR run does not show the 18\,min QPO. The broad feature is now replaced by two narrow peaks of amplitude not much different than flickering itself.

The power spectra of CVs show so called ``red noise'', thought to originate from the flickering through a shot noise-like process. This is demonstrated by the power-law decrease of the power with the frequency which is represented by a line in log-log scale. In this way, the power-law index $\gamma$ is used for the characterisation of flickering activity. It is believed that $\gamma=2$ corresponds to shots of infinite rise-time and exponential decay, while lower values to shots with different durations that follow some kind of distribution. For more information on ``red noise' and $\gamma$ interpretation, we refer to \citet{pap06} and references therein. ``Red noise'' is also present in our case and therefore the linear part, from 100\,$\rm c\rm d^{-1}$ on, was fitted by a least square fit in order to determine $\gamma$. It was found 1.41$\pm$0.03 for the OHL run, 1.17$\pm$0.03 for the KR one and 1.29$\pm$0.02 for all runs. It has to be mentioned that when \citet{pat01} applied the same procedure for their power spectra they found no QPO feature in the log-log power spectra (but only in the power spectra) and $\gamma$ had the value of 2. In our case the lower $\gamma$ suggests some kind of distribution in the shot-noise and this is reflected by the QPO or QPOs evident in our log-log power spectra.

\section{V516 Cyg}
Despite its early discovery as a variable star \citep{hof49} V516 Cyg was only noticed to be a DN by \citet{mei66}. It was decades later, that the first optical spectrum was obtained and V516 Cyg was confirmed to be a DN \citep{bru92}. The spectrum showed a strong blue continuum with Balmer and He lines in emission. The Balmer emission lines were superimposed on absorption troughs, characteristic of DN declining from (or rising to) an outburst. The first $\rm B$$\rm V$$\rm R$$\rm I$ photometric study during the decline from outburst was conducted by \citet{spo98}, who favoured an outside-in outburst. \citet{spo00} covered a whole outburst cycle. The long-term light curves can be found in the two aforementioned studies. The light curve of V516 Cyg during the whole outburst cycle pointed to the SS Cyg subclass of CVs.

\begin{figure}[!ht]
  \centerline{\includegraphics[width=1\textwidth]{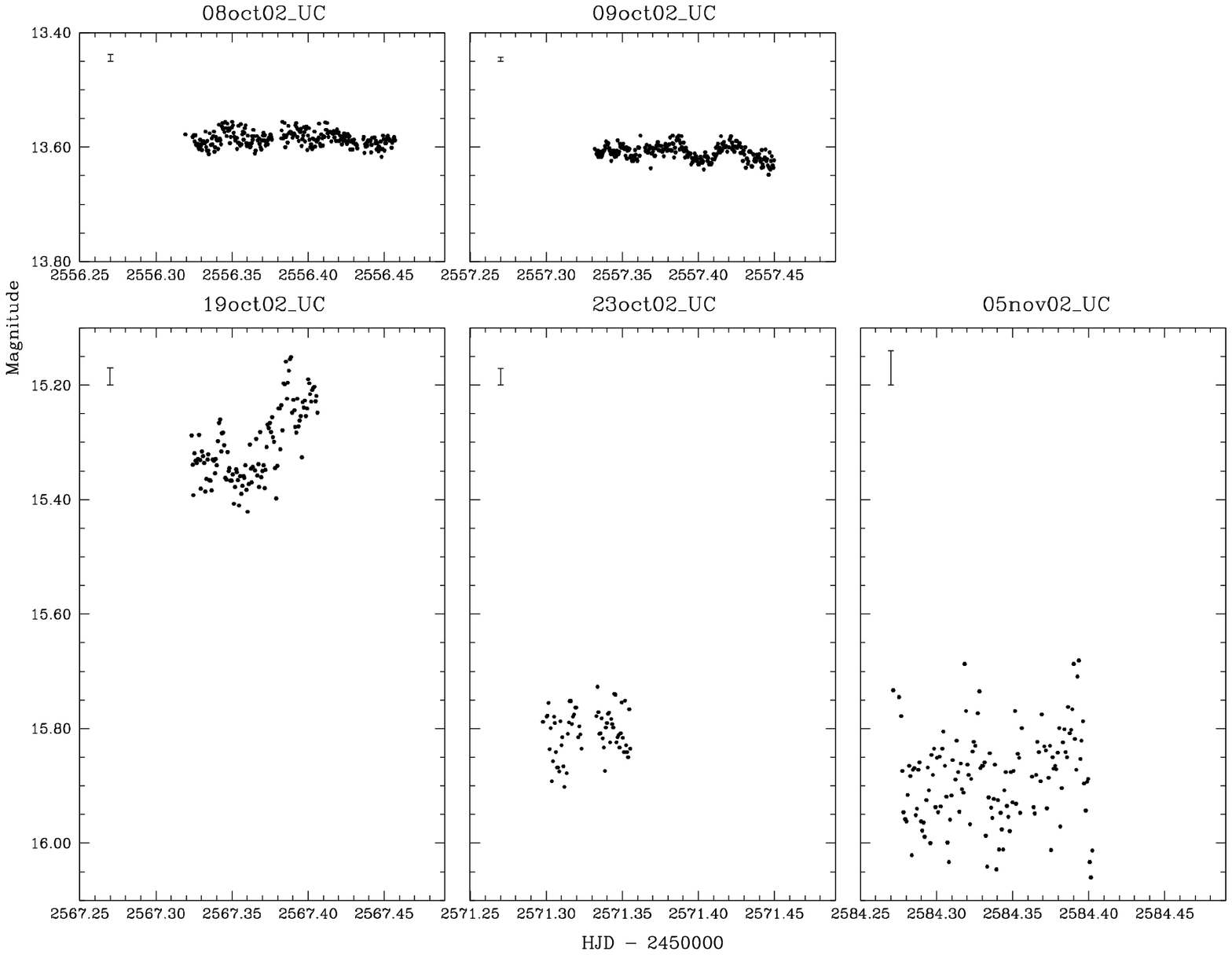}}
      \caption{V516 Cyg light curves. In the upper left corner of each light curve the error on a single measurement is shown.}
      \label{f_lc_V516Cyg}
    \end{figure}

V516 Cyg was observed at UC during 5 nights. No filter was used and the exposure time varied between 30 and 60\,sec. The light curves are shown in Fig.~\ref{f_lc_V516Cyg}. During the first two observing nights the system was at the peak of an outburst, confirmed by AAVSO long-term light curves. During the remaining nights the system was returning to quiescence, having reached it the last night. The magnitude difference between the two states was $\approx$2.5\,mag. 

Frequency analysis was applied to the dataset, as well as to the two subsets independently. However, no dominant periodicity was revealed. The system's decrease in brightness was followed by an increase in flickering activity. The amount of flickering was measured and found to be 0.0129$\pm$0.0004\,mag in outburst. Although still greater, the flickering activity during quiescence, in reality, is not much different than that of outburst since the error has incereased almost proportionally to the amount of flickering.

\section{GZ Cnc}
GZ Cnc (Tmz V034, RX J0915.8+0900) was discovered by \citet{tak98} and identified as a ROSAT source by \citet{bad98}. \citet{jia00} obtained the first optical spectrum which confirmed it as a CV. Some noticeable brightenings \citep{kat01} set the onset of its monitoring, and the detection of outbursts confirmed it as a DN \citep{kat01,kat02}. Photometry during the decline from outburst \citep{kat01} suggested that GZ Cnc should be a long-period DN, probably of the SS Cyg subtype. This was based on the facts that GZ Cnc showed no superhumps, had a low outburst amplitude and a slow rising at a rate similar to CVs of the SS Cyg subtype. Their dataset consists of 3 nights, the longest one spanning $\approx$3.3\,h. The long-term light curve of GZ Cnc was studied by \citet{kat02}. They claimed that it showed anomalous clusterings of outbursts in 2002, in contrast to earlier years and they associated this with the behaviour seen in some intermediate polars (IPs). This would result in a period greater than 3\,h. However, radial velocities of the H$\alpha$ emission line revealed the system's $\rm P_{\rm orb}$ of 2.118$\pm$0.007\,h \citep{tap03}. The same study also favoured the IP model due to the strength of the HeII line and the appearance of an absorption component during the rise to outburst. They summed up that even though there were strong indications that GZ Cnc could be an IP, these indications could by no means be conclusive. Both high-speed photometry and time-resolved spectroscopy would certainly clarify its classification.

\begin{figure}[!ht]
  \centerline{\includegraphics[width=1\textwidth]{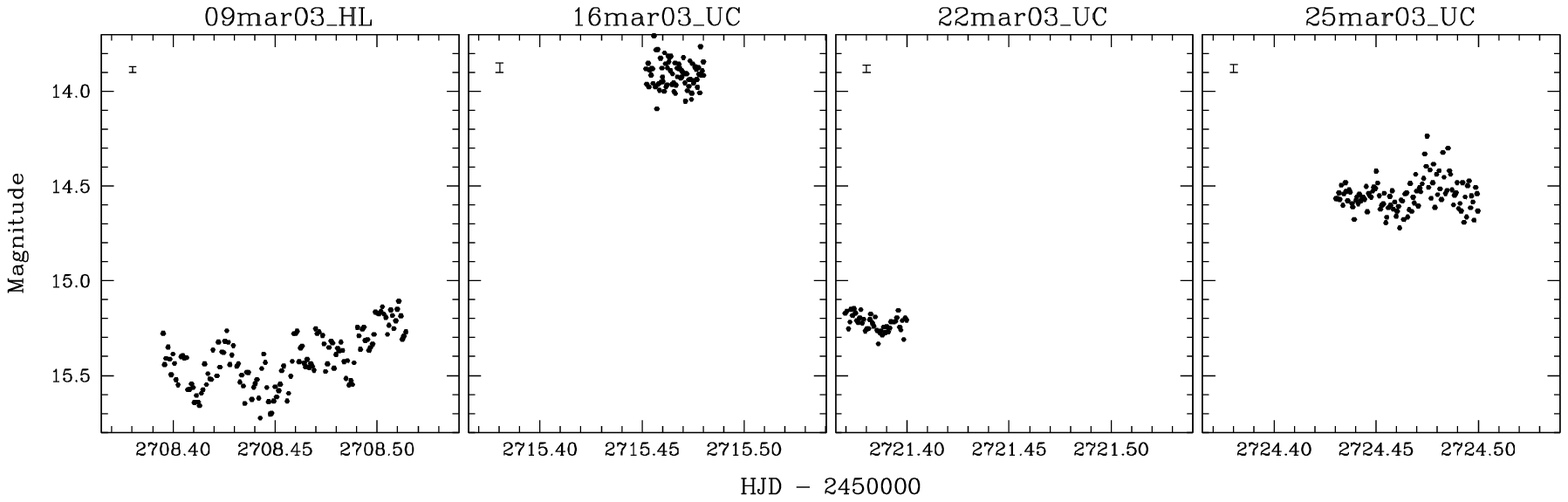}}
      \caption{GZ Cnc light curves. In the upper left corner of each light curve the error on a single measurement is indicated.}
      \label{f_lc_GZCnc}
    \end{figure}

This dataset consists of 1 night of observing at OHL and 3 at UC. All observations were unfiltered and the exposure time varied between 50 and 60\,sec. Since no well defined magnitudes exist, the ones from the USNO-A2 catalogue were used for computing the system's magnitude. Fig.~\ref{f_lc_GZCnc} shows the light curves. 

The system is caught during one of its frequent outbursts but unfortunately our coverage is not complete enough to reveal any periodicities. The only periodicity that comes up after excluding the two shortest nights is that of 6.5\,$\rm c \rm d^{-1}$ or 3.7\,h. However, the folding of the light curves on this periodicity and our maximum coverage per night which does not exceed 2.8\,h make us quite unsure of its reality. The amount of flickering was found as 0.08435$\pm$0.041\,mag.

We also searched for QPOs but none was evident, the small number of runs and especially of the long runs making a detection even more difficult. Even though both long runs revealed a rather broad feature around 25\,min, its amplitude was similar to that of the continuum making its reality doubtful.

\section{TY Psc}
The outbursting character of the SU UMa system TY Psc was first studied from AAVSO light curves \citep{szk84}. General and mean outburst characteristics can be found in Table~1 and Table~2 of the same study. A 1.7-h photometric modulation was reported by \citet{rob83}. UV and optical spectra revealed that TY Psc is a weak source in both cases \citep{szk85}. The optical spectra of the same study showed broad absorption dips around the emission lines in the blue region of the spectrum, thus indicating a substantial contribution of the WD. Infrared light curves \citep{szk88} showed a variation at 46\,min. This they identified as the ellipsoidal variation caused by the secondary star, a conclusion they supported by the following arguments. A private communication of theirs with J. Mattei revealed that the system showed superhumps with a period of 101\,min. This would mean that $\rm P_{\rm orb}$ would be shorter and 92\,min which is the double of the ellipsoidal variation would be a good fit. Also their $J-H$ values supported a large contribution of the secondary star. The same authors gave a rough estimation of the inclination at 55$^{\circ}$. Spectrophotometry by \citet{tho96} confirmed the strong WD contribution as noted before by \citet{szk85}, as well as a $P_{\rm orb}$ of 98.4\,min, from both radial velocities and photometry. Both the orbital modulation and double-peaked emission lines pointed to a rather high inclination, while the hot spot was also identified in the light curve. \citet{cia98} performed time-resolved near-infrared broadband photometry and concluded that the secondary star contributed less than 20\% of the infrared flux therefore opposing the idea of the ellipsoidal variation \citep{szk88}. Their light curve was quite similar to the optical one \citep{tho96}. This resemblance was interpreted as a result of the hot spot in the accretion disc dominating the emission and not as the secondary star contribution, which they found insufficient. Last, the superhump period of 101.9\,min was confirmed through observations during the 2000 superoutburst \citep{kun01}.

\begin{figure}[!ht]
  \centerline{\includegraphics[width=1\textwidth]{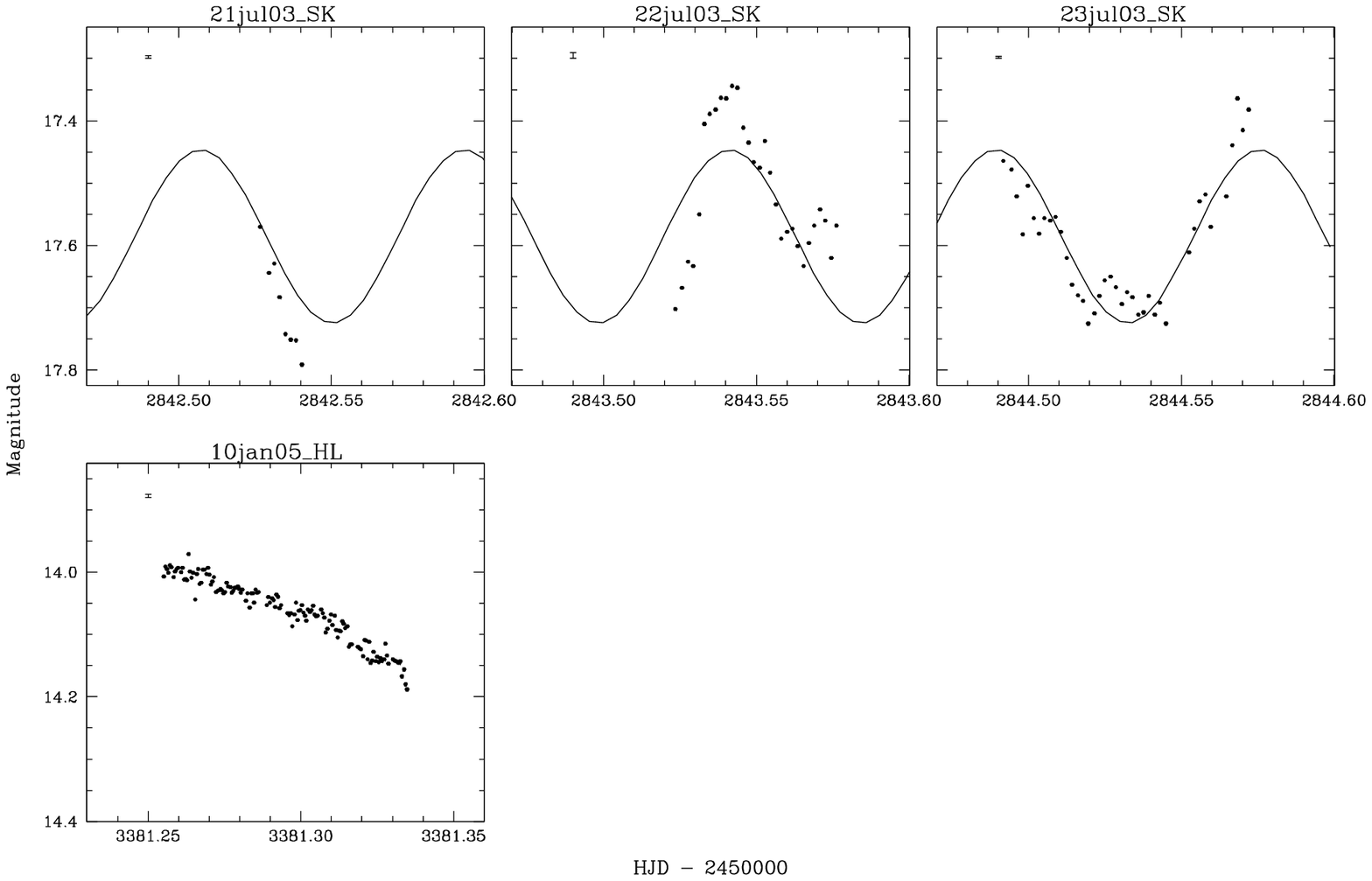}}
      \caption{TY Psc light curves. In the upper left corner of each light curve the error on a single measurement is indicated. The light curves on the top row correspond to quiescence, while the one on the bottom row to outburst.}
      \label{f_lc_TYPsc}
    \end{figure}

TY Psc was observed at both SK and OHL for 3 and 1 nights, respectively. The exposure time was 40--120\,sec and no filters were used. Fig.~\ref{f_lc_TYPsc} shows the light curve during the SK run that TY Psc was in quiescence (top row) and during the OHL run that it was in outburst (bottom row).

Frequency analysis was performed during quiescence (SK light curves) and a frequency of 11.60$\pm$0.02\,$\rm c\rm d^{-1}$ with an amplitude of 0.14$\pm$0.01\,mag was found. This frequency corresponds to an alias of the $P_{\rm orb}$ of 14.63\,$\rm c \rm d^{-1}$ and the power spectrum reveals the large number of aliases around our periodicity, including the $P_{\rm orb}$. Our periodicity of 2.01\,h is overplotted on the light curves of Fig.~\ref{f_lc_TYPsc}. For a check we also tried to fold the light curves on the $P_{\rm orb}$, something that gave subtle differences. Concerning the single night in outburst, we did not use it in the analysis. However, if we notice the corresponding light curve of  Fig.~\ref{f_lc_TYPsc} we can easily see that the $P_{\rm orb}$ would not fit. Such a thing could be caused by the fact that the system, as confirmed by AAVSO light curves, has just started its decline from the outburst peak. Such a fast decline ($\approx$3\,d for this system) could temporarily alter the morphology of the light curve.

After removing the periodicity from the SK light curves and the trend from the OHL one, we measured the amount of flickering. It was found 0.042$\pm$0.003\,mag and 0.012$\pm$ 0.004\,mag for the quiescence and outburst, respectively. The poor sampling of the data points in quiescence did not permit us to look for QPOs.

\section{ASAS J002511+1217.12}

ASAS J002511+1217.12 (1RXS J002510.8+121725) was discovered by \citet{poj02} and the All-Sky Automated Survey (ASAS). It was also detected by ROSAT \citep{vog99}. Upon an alert of a dramatic increase in brightness, astronomers started following the object for a 2-month period. The long term light curve, first presented by \citet{gol05}, revealed that ASAS J002511+1217.12 was an excellent WZ Sge candidate. The same light curves, accompanied by spectroscopic observations, were extensively studied by \citet{tem06} who classified it as a new WZ Sge star. WZ Sge stars have large amplitude infrequent outbursts, orbital periods around 2\,h, show superhumps, and a slow decline from maximum light to quiescence during which echo outbursts are occasionally formed. Echo outbursts resemble DN outbursts and appear directly after the main outburst. They are of about similar amplitude and recurrence time while, after their complete cessation the system slowly returns to quiescence \citep{pat02}. ASAS J002511+1217.12 does not lack any of these characteristics and the results of \cite{tem06} will now be briefly presented.

The system's light curve spans for 3 months (mid-September 2004 -- mid-December 2004) and during that period the system declined by $\approx$7\,mag. The early abrupt decline was followed by one echo outburst and the decay time to quiescence lasted $\approx$90 days. Time-series CCD photometry during the early outburst showed clear superhumps at 81.9\,min shifting around this value as the system faded. During the quiescent state, just prior the echo outburst, variations at 81.6\,min were detected. During the echo outburst there appeared to be weak variations of the same period as the original superhumps. A few nights of photometry during the late outburst (after the echo outburst) showed signals at periods nearly the same as the superhumps, however it was not as easy to distinguish due to the increased photometric noise. Time-series spectroscopic observations, well into the object's decline, yielded a $\rm P_{\rm orb}$ of 82$\pm$5\,min, albeit not too precise. Narrow and broad components in the emission-line spectra indicated the presence of multiple emission regions.

\begin{figure}[!ht]
  \centerline{\includegraphics[width=1\textwidth]{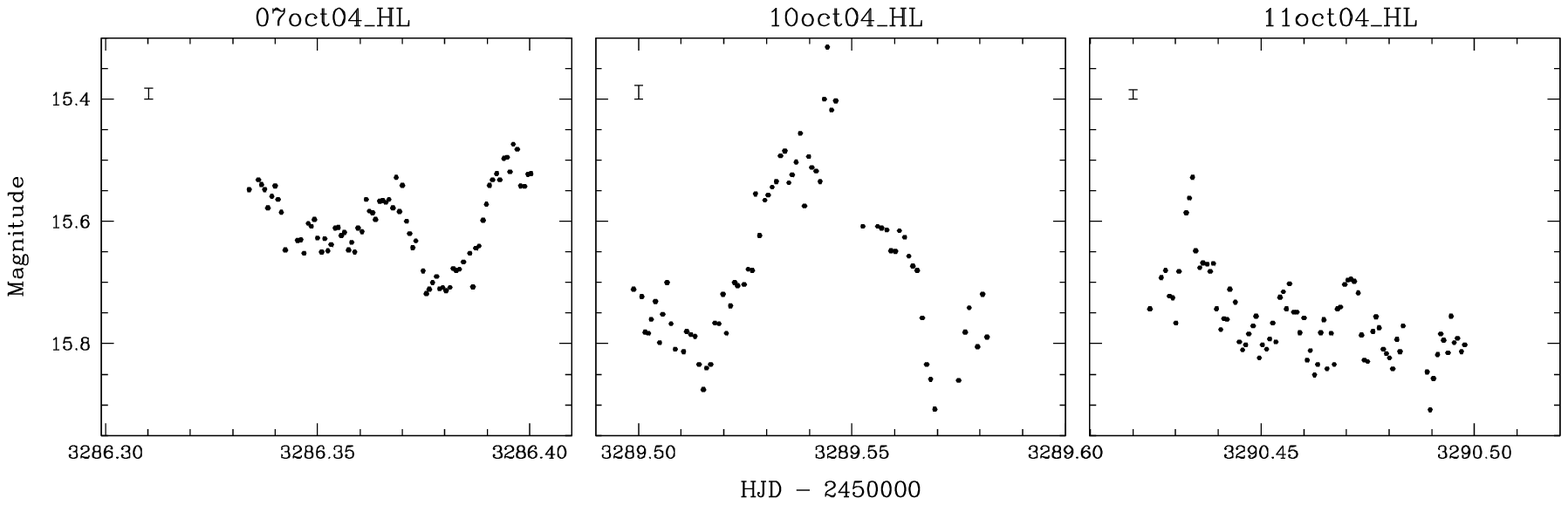}}
      \caption{ASAS J002511+1217.12 light curves. In the upper left corner of each light curve the error on a single measurement is indicated.}
      \label{f_lc_Psc}
    \end{figure}

ASAS J002511+1217.12 was observed during 3 nights at OHL, right after the 2004 echo outburst. The exposure varied between 15 and 60\,sec, and the light curves are shown in Fig.~\ref{f_lc_Psc}. 

Our observations lying after the end of the echo outburst show similar results with \cite{tem06}. We find a frequency of 13.42$\pm$0.12\,$\rm cd^{-1}$ with an amplitude of 0.08$\pm$0.01\,mag and an alias of the superhump modulation. However the folding is not satisfactory. The amount of flickering was found to be 0.070$\pm$0.017.

\section{V1315Aql}
V1315 Aql (KPD 1911+1212, SVS 8130) was discovered and suspected as a variable star by \citet{met61}. Later on, \citet{dow86} based on a spectrophotometric study classified this object as an eclipsing CV, with an eclipse depth up to 1.8\,mag in $V$. The $\rm P_{\rm orb}$ was found to be 3.35\,h. Apart from Balmer and HeI emission lines the spectra showed strong high-excitation lines. Even though the HeII $\lambda$4686 was totally eclipsed, the Balmer lines were not. A prominent absorption component was visible at phase 0.5. 

These spectral characteristics, combined with the occurrence of single-peaked emission lines despite the system's high inclination \citep{szk90}, uncommon at that time, were the onset of the system's follow up and the emergence of a new sub-class of NL CVs, the SW Sex stars. A detailed spectroscopic study by \citet{dhi91} showed that the spectra of V1315 Aql have: (1) narrow single-peaked Balmer and HeI lines exhibiting a double-peaked structure near phase 0.5 and (2) high-excitation lines, remaining single-peaked throughout the orbit and being totally eclipsed in contrast to Balmer and HeI lines. They presented the first Doppler maps which showed a contaminating emission around phase 0.75 and no disc emission. This, they connected to the large phase shifts seen in the radial-velocity measurements of all emission lines. Even though the next Doppler maps presented by \citet{kai94} appeared quite similar, there was one disagreement concerning the origin of the HeII peak emission. Although \citet{dhi91} did not find it to coincide with the expected location of the WD, \citet{kai94} did. In this way they drew the conclusion that HeII is mostly emitted from a region very close to the WD and is therefore a good indicator of the WD orbital motion.
 
Further spectroscopy \citep{fri88,smi93} showed a strong absorption feature at the $\rm OI$ triplet $\lambda$7773. A detailed study by \citet{smi93} concluded that this feature must arise from some absorbing material that is distributed in a non-axisymmetric manner. This asymmetry seemed to be confined only to $\rm OI$ . Infrared spectroscopy showed no evidence for the secondary star \citep{dhi00}. For an analytic presentation of a model that could describe the characteristics and peculiarities of V1315 Aql and therefore all SW Sex stars, as well as previous attempts on the same subject, see \citet{hel96} and references therein. In short, the model combines the occurrence of both disc overflow and a strong wind, which might result from high mass-transfer rate.

\begin{figure}[!ht]
  \centerline{\includegraphics[width=1\textwidth]{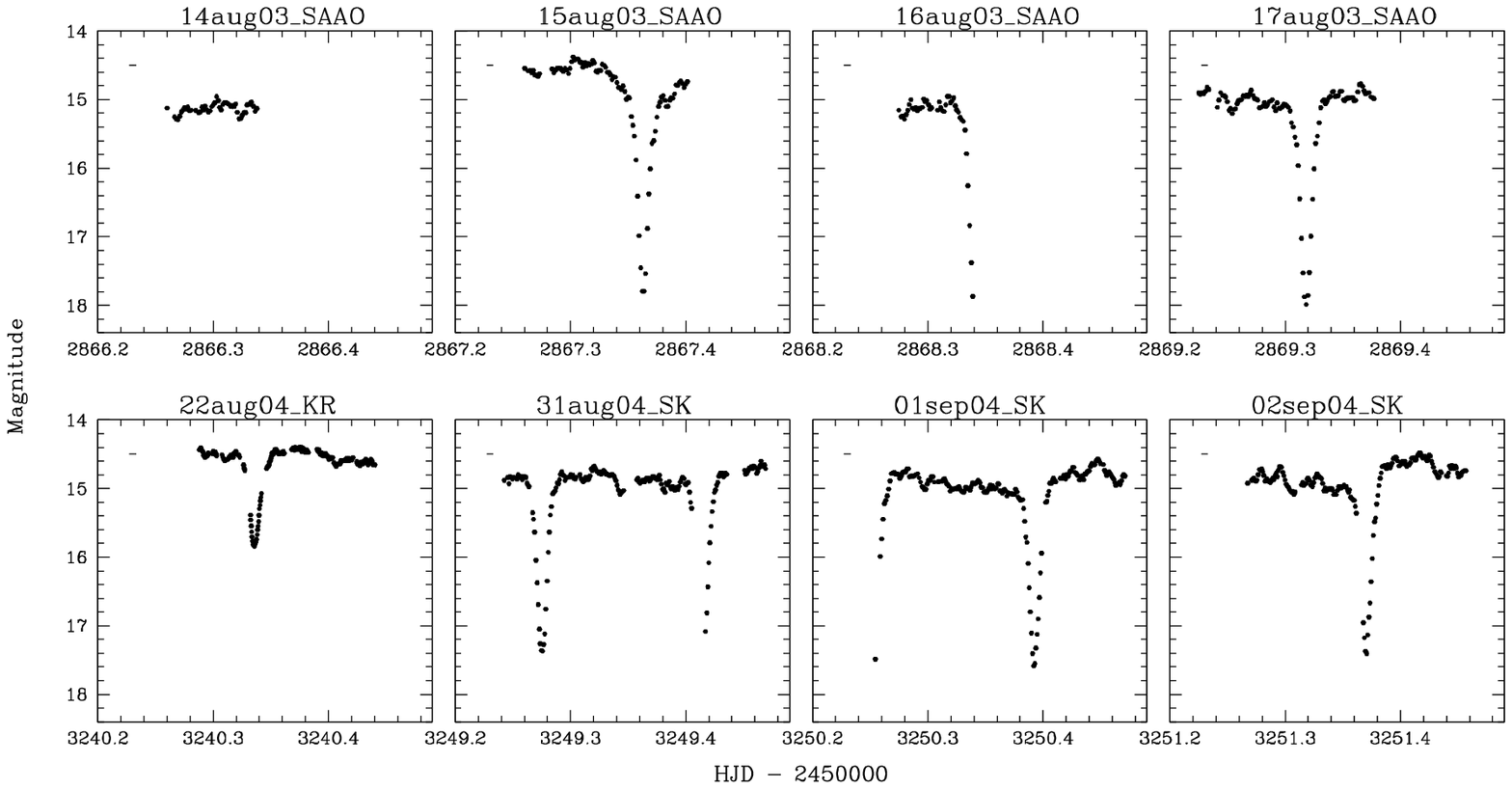}}
      \caption{V1315 Aql light curves. In the upper left corner of each light curve the error on a single measurement is indicated.}
      \label{f_lc_V1315}
    \end{figure}

V1315 Aql was observed at SAAO (4 nights), at SK (3 nights) and at KR (1 night). At SK the Johnson-$R$ filter was used, while the rest of the runs were unfiltered. The exposure time was 40-60\,sec. 
The light curves are shown in Fig.~\ref{f_lc_V1315}.

V1315 Aql was observed during two periods, August 2003 and August 2004. In total, the system was caught during nine eclipses but, only six of them were fully covered and are therefore used for further analysis. The timings of mid-eclipse, determined by fitting a gaussian to the lower half of the eclipses, are shown in Table~\ref{t_minima}. The same table also includes all previously reported minima and their corresponding cycle number $E$. The refined ephemeris resulting from all available eclipse timings is:

\vskip 0.5cm 
\begin{tabular}{ccrcr}
$T_{\rm mid-\rm ecl}[\rm HJD]$ & $=$ & $2445902.7007$& $+$& $0^{\rm d}.139689959\,E$\\
                               & & $\pm0.0001$ && $\pm0.000000005\,\,\,\,\,\,$ \\
\end{tabular}
\vskip 0.5cm

\begin{figure}[!ht]
  \centerline{\includegraphics[width=0.8\textwidth]{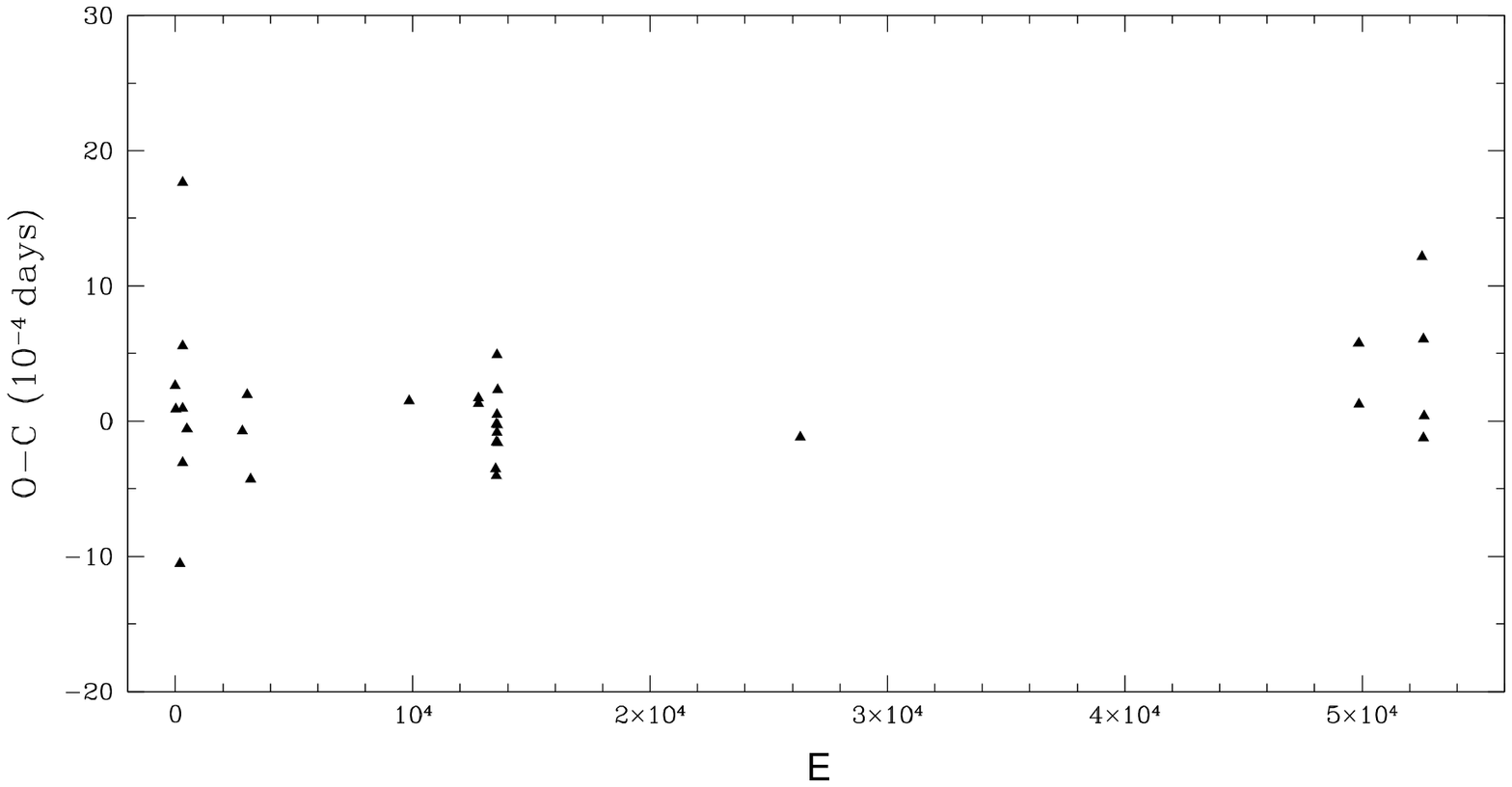}}
      \caption{$O-C$ diagram for the timings of mid-eclipse of all available data (see Table~\ref{t_minima}).}
      \label{f_oc}
    \end{figure}

Fig.~\ref{f_oc} shows the $O-C$ diagram of all available minima. $O$ represents the observed times of minima and $C$ the calculated ones according to the ephemeris. In agreement to all previous studies, no evidence of any kind of curvature or trend is found in the $O-C$ diagram. This means that in the now even longer baseline of 20 years there is no indication of a change in the system's period.
    
A look at Fig.~\ref{f_oc} shows that there is a large scatter of data points at the first and last block of data. However, concerning the last block belonging to our observations, the variations we are looking at are of the order of 40-50 sec, i.e. of the size of the exposure times and therefore they shouldn't be of any significance. Unfortunately, no errors are given on any of the previous studies and so we can not reach any conclusion about the even larger scatter of the first dataset.

\begin{table}[!ht]
  \caption{Eclipse timings of V1315 Aql}
  \label{t_minima}
  \centering
  \begin{tabular}{|lll|}
    \hline
    $\rm T_{\rm mid-\rm ecl}$& E  & Source \\
    {\small (HJD-2445000)}&        & \\
    \hline 
     902.84065   & 1	&    \citet{dow86}	\\
     906.75180   & 29	&    \citet{dow86}	\\
     928.82167   & 187	&    \citet{ann86}	\\
     944.74793   & 301	&    \citet{dow86}	\\
     944.88716   & 302	&    \citet{dow86}	\\
     945.72697   & 308	&    \citet{ann86}	\\
     945.86459   & 309	&    \citet{dow86}	\\
     971.70748   & 494	&    \citet{dow86}	\\
    1295.78815   & 2814	&    \citet{ann86}	\\
    1324.70424   & 3021	&    \citet{ann86}	\\
    1344.67928   & 3164	&    \citet{ann86}	\\
    2279.62471   & 9857	&    \citet{dhi91}	\\
    2686.54156   & 12770&    \citet{dhi91}	\\
    2686.68121   & 12771&    \citet{dhi91}	\\
    2788.37501   & 13499&    \citet{dhi91}	\\
    2790.47069   & 13514&    \citet{dhi91}	\\
    2791.44839   & 13521&    \citet{dhi91}	\\
    2793.40380   & 13535&    \citet{dhi91}	\\
    2794.52177   & 13543&    \citet{dhi91}	\\
    2795.49947   & 13550&    \citet{dhi91}	\\
    2796.47787   & 13557&    \citet{dhi91}	\\
    2798.43327   & 13571&    \citet{dhi91}	\\
    2800.38867   & 13585&    \citet{dhi91}	\\
    2801.36637   & 13592&    \citet{dhi91}	\\
    4579.6195    & 26322&    \citet{hel96}	\\
    7867.36291   & 49858   &   This study	  \\
    7869.31812   & 49872  &   This study	\\
    8240.33572   & 52528  &   This study	  \\
    8249.27527   & 52592  &   This study	  \\
    8250.39206   & 52600  &   This study	\\
    8251.37005   & 52607  &   This study	\\
    \hline       	       
 \end{tabular}   	       
\end{table}

The application of frequency analysis, after masking the eclipses, gives a frequency of $5.14627 \pm 3 \cdot 10^{-5}\,\rm c \rm d^{-1}$ an alias of the $P_{\rm orb}$. We also searched for long-term periodicities, but none was evident. The same applies for QPOs, where none was convincing enough to be adopted. The amount of flickering was found to be 0.117$\pm$0.034\,mag.

\begin{figure}[!ht]
  \centerline{\includegraphics[width=1\textwidth]{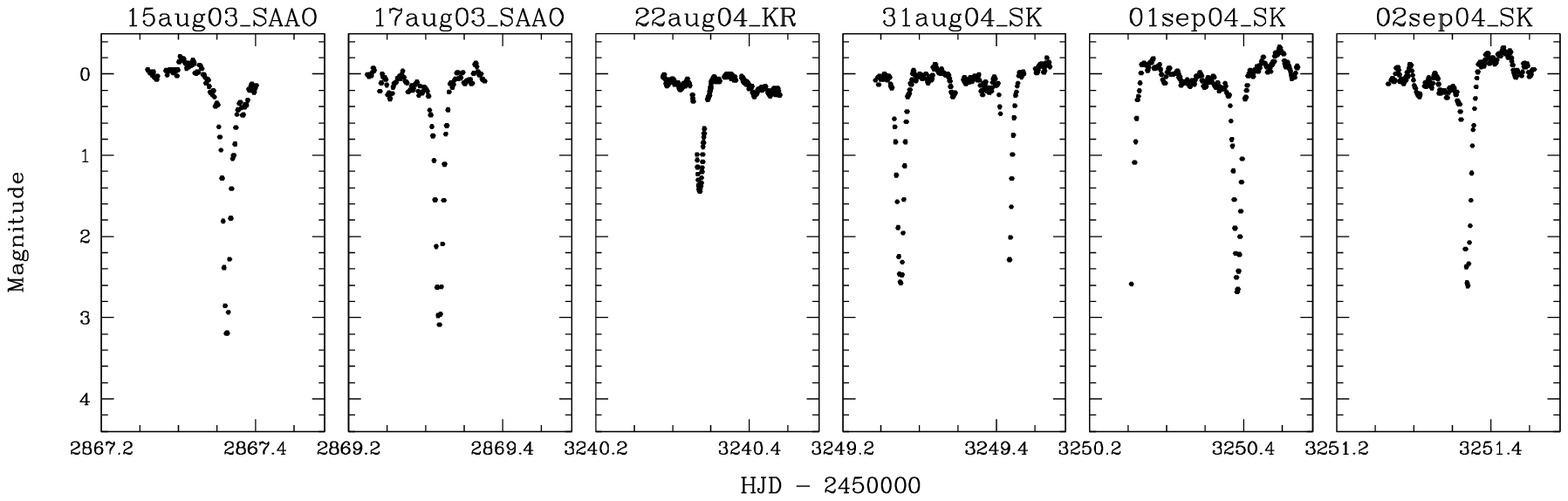}}
      \caption{V1315 Aql eclipse light curves with the out-of-eclipse magnitude shifted to zero. The eclipse depth varies between 1.5 and 3.2 in the two extreme nights.}
      \label{f_ecl}
    \end{figure}

As previously mentioned, V1315 Aql undergoes eclipses of around 1.8\,mag in $V$. In our case the eclipse depth is found around 3.2 and 2.6\,mag in the unfiltered 2003 and $R$-band 2004 runs, respectively. The only exception is the eclipse depth of the unfiltered night of $22^{\rm nd}$ of August 2004. Here the eclipse depth is remarkably smaller at 1.5\,mag, half and less than half of the rest of the nights. In Fig.~\ref{f_ecl} the out-of-eclipse magnitude has been shifted to zero and the variable eclipse depth is readily estimated. Variable eclipse depth has also been reported for DW UMa \citep{sta04} and PX And \citep{sta02}. In DW UMa it was attributed to the change between high and low state of the accretion disc radius and therefore the surface of the eclipsed accretion disc. However, in our study the unique low-depth eclipse makes it hard to investigate further and try to draw firm conclusions. Even if we tried to associate the eclipse depth with the system luminosity we can directly see that between the two extreme nights of 15 August 2003 and 22 August 2004 where the eclipse depth is $\approx$ 1.5 and 3.2\,mag, the out-of-eclipse magnitude is the same (see Fig.~\ref{f_lc_V1315}). In PX And the eclipse-depth was suggested to be modulated with the precession period of the accretion disc. However, in the case of V1315~Aql no longer-term modulation has been suggested so far. 
\section{LN UMa}

LN UMa (UMa7, PG 1000+667) was discovered in the Palomar-Green survey \citep{gre86} and, based on its spectral characteristics, was classified as a NL CV \citep{rin93}. \citet{rin93} found a period of 4.06\,h. However, the spectrophotometric study of \citet{hil98} revealed a refined $\rm P_{\rm orb}$ of 3.47\,h through H$\beta$ radial velocities. Their multi-year photometry (1400 days covering 1 point every 4.7 days) showed two drops of 3 magnitudes \citep{hil98,hon04}. Given the $\rm P_{\rm orb}$ and the low states, LN UMa was identified as a VY Scl star. Not much later, a spectroscopic search for new SW Sex stars revealed that LN UMa is also part of this interesting subclass of CVs \citep{rod07}.

\begin{figure}[!ht]
  \centerline{\includegraphics[width=1\textwidth]{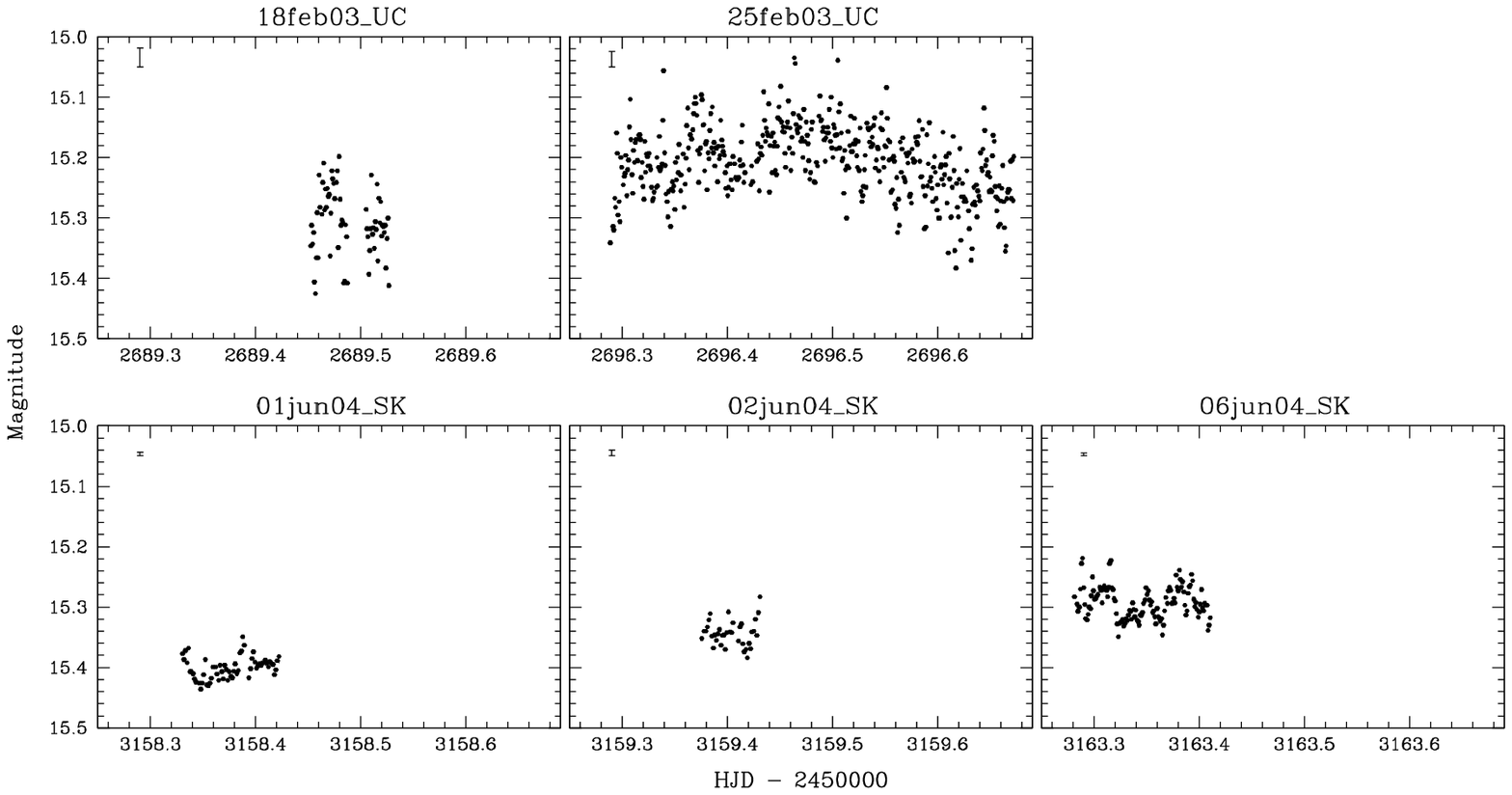}}
      \caption{LN UMa light curves. In the upper left corner of each light curve the error on as single measurement is shown.}
      \label{f_lc_UMa7}
    \end{figure}

\begin{figure}[!ht]
  \centerline{\includegraphics[width=1\textwidth]{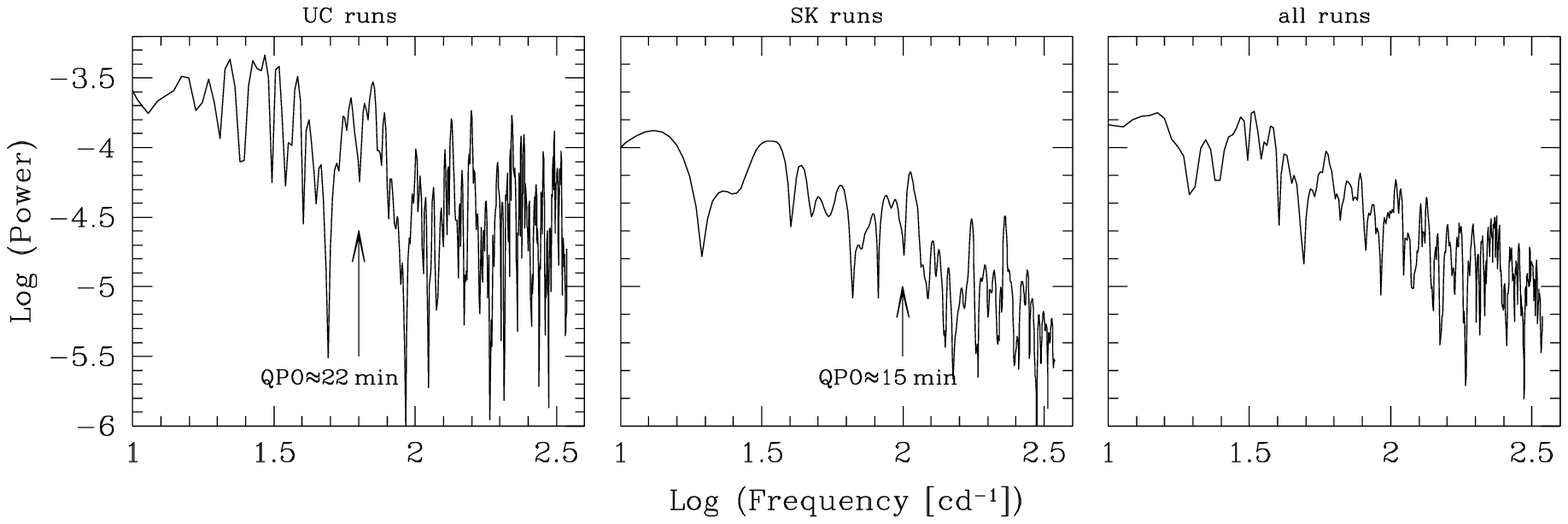}}
      \caption{From left to right: LN UMa average power spectra for the UC 2003 run, SK 2004 runs and all runs, respectively. The arrows show the the frequency of the possible QPOs. }
      \label{f_qpo_UMa7}
    \end{figure}

Our dataset consists of five observing runs, three at SK and two at UC. At SK we observed using the Johnson-$R$ filter while the UC run was unfiltered. The exposure time varied between 60 and 80\,sec. The light curves are shown in Fig.~\ref{f_lc_UMa7}.

After checking the long-term light curves of our system from AAVSO it is clear that the UC run is placed during the normal high state, but this is not evident for the SK run. However, the trend of the AAVSO light curves points to the normal high state for the SK run as well. As seen in Fig.~\ref{f_lc_UMa7}, the atmospheric conditions for the SK run are much better than the UC one, with the error being greater in the UC run. This affects the flickering of LN UMa which is found 0.0554\,mag for the UC and 0.0218\,mag for the SK run. The last value is the one that should be trusted.  

Period analysis to all of our dataset as well as the individual runs did not reveal the spectroscopic $P_{\rm orb}$ or any other modulations. The high amplitude flickering of the longest run, covering more than two orbits, could easily mask any underlying periodicity. However, a possible QPO resolving into 2 components was found in the UC and SK run, as shown in  Fig.~\ref{f_qpo_UMa7}. In the UC run it is found around 22\,min, while in the SK run around 15\,min. The sampling of the light curves was such that we could not reach frequencies higher than 345\,$\rm c \rm d^{-1}$. Therefore, no fits of the linear parts were made because we could be driven to inaccurate values of $\gamma$.

\section {Summary}
The results of this study can be summarised as follows:
\begin{description}
\item [EI UMa] 
For the first time we confirm through photometry the spectroscopic $P_{\rm orb}$ of 6.4\,h. Its rather low amplitude of 0.021$\pm$0.002\,mag made its direct detection in 2002 more difficult and was probably the reason for its non-detection so far. Moreover, in 2002 the system exhibits a gradual increase in brightness reaching a maximum of 0.6\,mag and a high amplitude unstable modulation. The increase in brightness is followed by a decrease in the amount of flickering.
\item [V844 Her] 
The spectroscopically defined $P_{\rm orb}$ could not be detected. The same
applies for the $P_{\rm sh}^+$ but this was expected since the system was in quiescence. Unluckily, AAVSO long-term light curves confirm that our two observing run lie just before the onset and just after the end of a superoutburst.
\item [V751 Cyg] 
We have detected the two previously reported photometric signals $P_{\rm sh}^-$
and $P_{\rm wob}$, albeit only their aliases and harmonics in some cases. \citet{pat01} reported a possible QPO at $\approx20$\,min and we have confirmed it at $\approx$18\,min. This QPO was present at the OHL run but not at the KR one. In contrast to \citet{pat01} who found $\gamma=2$ we found it at 1.29$\pm$0.02 which implies that the flickering activity has changed and in our case the shot-noise follows some kind of distribution. 
\item [V516 Cyg] 
The system was observed at the peak of an outburst and during its decline, until it reached its quiescent state. Between the two extreme states a magnitude difference of 2.5\,mag was observed. No dominant periodicity could be detected. The decrease in brightness was followed by an increase in flickering activity, expected since the accretion disc returns to its normal smaller size and the relative contribution of the BS, major cause of the flickering activity, increases. 
\item [GZ Cnc] 
No periodicity could be detected during one of the system's frequent outbursts. The only possible modulation is one of 3.7\,h. However, the unsatisfying folding of our data on this modulation as well as the fact that it exceeds the duration of our longest observing night, does not add any credit.
\item [TY Psc] 
This system was observed during quiescence and during the initiation of decline from the peak of an outburst.We detected the $P_{\rm orb}$  during quiescence but not in outburst. The amount of flickering decreased with increasing brightness. 
\item [ASAS J002511+1217.12] 
Our results resemble those of \citet{tem06} which were also obtained after the echo outburst. The light curves exhibit high amplitude flickering and show two possible signals: at 13.42\,$\rm c\rm d^{-1}$ and an alias of the superhump modulation.
\item [V1315 Aql] 
Our observations covered nine more eclipses which combined with the previously reported mid-eclipse timings provide a refined orbital ephemeris. The $O-C$ shows that during the 20-year baseline there is no trend or curvature and therefore no change in the system's period. The eclipse depth varies between 2.6 and 3.2\,mag with the exception of one night where it has a remarkably smaller value of 1.5\,mag. For the moment no association between the brightness level and the eclipse depth can be made because the two nights with the extreme values of the eclipse depth (1.5 and 3.2\,mag) share the same out-of-eclipse magnitude.
\item [LN UMa] 
Our observing set is probably placed during the system's normal high state and does not reveal the spectroscopic $P_{\rm orb}$ or any other modulation. The only exception is a possible QPO around 22\,min in the UC run and around 15\,min in the SK run. In both cases it resolves into 2 components.
\end{description}

\section{The data}

\subsection{Tables}
The files table1.dat--table4.dat are ASCII files of the same tables as presented in this manuscript. 

\subsection{Light curve data}
The files fig2.dat--fig15.dat (the same numbering as in the manuscript is followed) are ASCII files with the light curves of EI UMa, V844 Her, V751 Cyg, V516 Cyg, GZ Cnc, TY Psc, ASAS J002511+1217.12, V1315 Aql and LN UMa, respectively. They give the HJD and differential magnitude of each CV.

Moreover, the instrumental magnitudes (i.e. those before the application of differential photometry) of each CV and its comparison stars are given in ASCII files with HJD, magnitude and magnitude error. Those files are table5.dat--table13.dat  for EI UMa, V844 Her, V751 Cyg, V516 Cyg, GZ Cnc, TY Psc, ASAS J002511+1217.12, V1315 Aql and LN UMa, respectively. The fisrt column gives the HJD, the second and third give the instrumental magnitude and corresponding error of the CV, the fourth and fifth give the instrumental magnitude and corresponding error of the first comparison star, the sixth and seventh give the instrumental magnitude and corresponding error of the second comparison star and so on.

\subsection{Calibrated frames}
The directories EIUMa\_frames, V844Her\_frames, GZCnc\_frames, V1315Aql\_frames, V516 Cyg\_frames, LNUMa\_frames, V751Cyg\_frames, TYPsc\_frames and ASAS\_frames contain the calibrated frames of each target as gzipped FITS files. The naming convention in each CV directory is \{UTdate\}\_\{sequence number\}.fits.gz . UT date represents the beginning of each observing night.

\newpage
\section*{Acknowledgments}

The first author carried out the bulk of the observations. \\

\noindent 
We are grateful for the generous allocations of time to	Prof. Klaus Reif and Prof. Wilhelm Seggewiss at the Observatorium Hoher List, to Dr. Iosif Papadakis, and Dr. Pablo Reig at Skinakas Observatory. Skinakas Observatory is a collaborative project of the University of Crete, the Foundation for Research and Technology-Hellas, and the Max--Planck--Institut f\"ur extraterrestrische Physik. This paper uses observations made at the South African Astronomical Observatory (SAAO). We wish to thank H.W. Duerbeck for comments that greatly improved the quality of the manuscript.  This work has been partly supported by ``IAP P5/36'' Interuniversity Attraction Poles Programme of the Belgian Federal Office for Scientific, Technical, and Cultural Affairs. C.P. gratefully acknowledges a doctoral research  grant by the Belgian Federal Science Policy Office (Belspo).


\end{document}